\documentclass[12pt]{article}
\usepackage{cite}
\usepackage{times}
\usepackage{fullpage}
\usepackage{ifthen}
\usepackage{url}
\newcommand{\spacing}[1]{\renewcommand{\baselinestretch}{#1}\large\normalsize}
\spacing{1}

\makeatletter \DeclareRobustCommand\cite{\unskip \@ifnextchar [{\@tempswatrue\@citex}{\@tempswafalse\@citex[]}} \def\@cite#1#2{$^{\hbox{\scriptsize{#1\if[@tempswa , #2\fi}}}$} \makeatother

\newenvironment{affiliations}{%
    \setcounter{enumi}{1}%
    \setlength{\parindent}{0in}%
    \slshape\sloppy%
    \begin{list}{\upshape$^{\arabic{enumi}}$}{%
        \usecounter{enumi}%
        \setlength{\leftmargin}{0in}%
        \setlength{\topsep}{0in}%
        \setlength{\labelsep}{0in}%
        \setlength{\labelwidth}{0in}%
        \setlength{\listparindent}{0in}%
        \setlength{\itemsep}{0ex}%
        \setlength{\parsep}{0in}%
        }
    }{\end{list}\par\vspace{12pt}}

\renewenvironment{abstract}{%
    \setlength{\parindent}{0in}%
    \setlength{\parskip}{0in}%
    \bfseries%
    }{\par\vspace{-6pt}}

\newcommand\aap{Astronomy \& Astrophysics}
\newcommand\apj{The Astrophysical Journal}
\newcommand\apjl{The Astrophysical Journal Letters}
\newcommand\apjs{The Astrophysical Journal Supplement}
\newcommand\aj{The Astronomical Journal}

\newcommand\pasj{Publications of the Astronomical Society of Japan}
\newcommand\physrep{Physics Reports}
\newcommand\nat{Nature}
\newcommand\mnras{Monthly Notices of the Royal Astronomical Society}

\newcommand\procspie{SPIE proceedings}

\usepackage{color,amsmath,amssymb}
\usepackage{graphicx,units,wasysym,hyperref}

\usepackage{lineno}
\usepackage{caption}

\usepackage[normalem]{ulem}

\usepackage{color}

\makeatletter
\def\@maketitle{
\begin{flushleft}
{\LARGE\bfseries\textsf{\@title} \par}
\end{flushleft}
\begin{flushleft}
{\@author \par}
\end{flushleft}
}

\title{High-entropy ejecta plumes in Cassiopeia A from neutrino-driven convection}

\author{Toshiki Sato,\!$^{1,2,3\ast}$ 
        Keiichi Maeda,\!$^{4}$
        Shigehiro Nagataki,\!$^{5,6}$
        Takashi Yoshida,\!$^{7}$
        Brian Grefenstette,\!$^{8}$
        Brian J. Williams,\!$^{2}$
        Hideyuki Umeda,\!$^{7}$
        Masaomi Ono,\!$^{5,6}$
        John P. Hughes \!$^{9}$
}
\begin{document}

\date{}

\maketitle

\begin{affiliations}
  \item {RIKEN, 2-1 Hirosawa, Wako, Saitama 351-0198, Japan}
  \item {NASA, Goddard Space Flight Center, 8800 Greenbelt Road, Greenbelt, MD 20771, USA}
  \item {Department of Physics, University of Maryland Baltimore County, 1000 Hilltop Circle, Baltimore, MD 21250, USA}
  \item{Department of Astronomy, Kyoto University, Kitashirakawa-Oiwake-cho, Sakyo-ku, Kyoto 606-8502, Japan}
  \item{Astrophysical Big Bang Laboratory (ABBL), RIKEN Cluster for Pioneering Research, 2-1 Hirosawa, Wako, Saitama 351-0198, Japan}
  \item{RIKEN Interdisciplinary Theoretical and Mathematical Science Program (iTHEMS), 2-1 Hirosawa, Wako, Saitama 351-0198, Japan}
  \item{Department of Astronomy, Graduate School of Science, University of Tokyo, 7-3-1 Hongo, Bunkyo-ku, Tokyo 113-0033, Japan}
  \item{Cahill Center for Astrophysics, 1216 E. California Boulevard, California Institute of Technology, Pasadena, CA 91125, USA}
  \item{Department of Physics and Astronomy, Rutgers University, 136 Frelinghuysen Road, Piscataway, NJ 08854-8019, USA}

$^{\ast}$e-mail: toshiki.sato@riken.jp, s.toshiki629@gmail.com
  \end{affiliations}
%%%%%%%%%%%%%%%%%%%%%%%%%%%%%%%%%%%%%%%%%%%%%%%%%%%%%%%%%%%%%%%%%%%%%%%%%%%%%%%

\begin{abstract}
Recent multi-dimensional simulations suggest that high-entropy buoyant plumes help massive stars to explode\cite{2016ARNPS..66..341J,2020MNRAS.491.2715B}. Outwardly protruding iron-rich fingers in the galactic supernova remnant
Cassiopeia A\cite{2000ApJ...528L.109H,2003ApJ...597..362H} are uniquely suggestive of this picture. Detecting signatures of specific elements synthesized in the high-entropy nuclear burning regime (i.e., $\alpha$-rich freeze out) would be among the strongest substantiating evidence.
Here we report the discovery of such elements, stable Ti and Cr, at a confidence level greater than 5$\sigma$ in the shocked high-velocity iron-rich ejecta of Cassiopeia A. We found the observed Ti/Fe and Cr/Fe mass ratios require $\alpha$-rich freeze out, providing the first observational demonstration for the existence of high-entropy ejecta plumes that boosted the shock wave at explosion. The metal composition of the plumes agrees well with predictions for strongly neutrino-processed proton-rich ejecta\cite{2006A&A...447.1049B,2018ApJ...852...40W,2020MNRAS.491.2715B}. These results support the operation of the convective supernova engine via neutrino heating in the supernova that produced Cassiopeia A.
\end{abstract}

The explosion mechanism leading to core-collapse supernovae (CC SNe) is a long-standing problem in astrophysics. The neutrino-driven explosion\cite{1985ApJ...295...14B} is a current viable mechanism that still needs to be tested by observation. In this mechanism, the outward shock wave formed by core bounce at the proto-neutron star's surface is boosted by strong neutrino heating. It has been recognized that convective overturn in the neutrino heating region is essential for initiating explosions\cite{1995ApJ...450..830B,2012PTEP.2012aA309J}. 
The buoyant high-entropy (low density and high temperature) bubbles then push the shock farther out. Some of the biggest high-entropy bubbles can develop into Rayleigh-Taylor instabilities on a larger scale at late stages of the explosion and penetrate into the hydrogen envelope with large velocities creating iron-dominated plumes\cite{2003A&A...408..621K,2010ApJ...714.1371H,2017ApJ...842...13W}. The presence of such high-entropy iron-dominated ejecta in SN remnants will allow us to probe details of the convective supernova engine. 

\begin{figure}
\internallinenumbers % required by nature
\resizebox{\hsize}{!}{\includegraphics[bb=0 0 1755 1486]{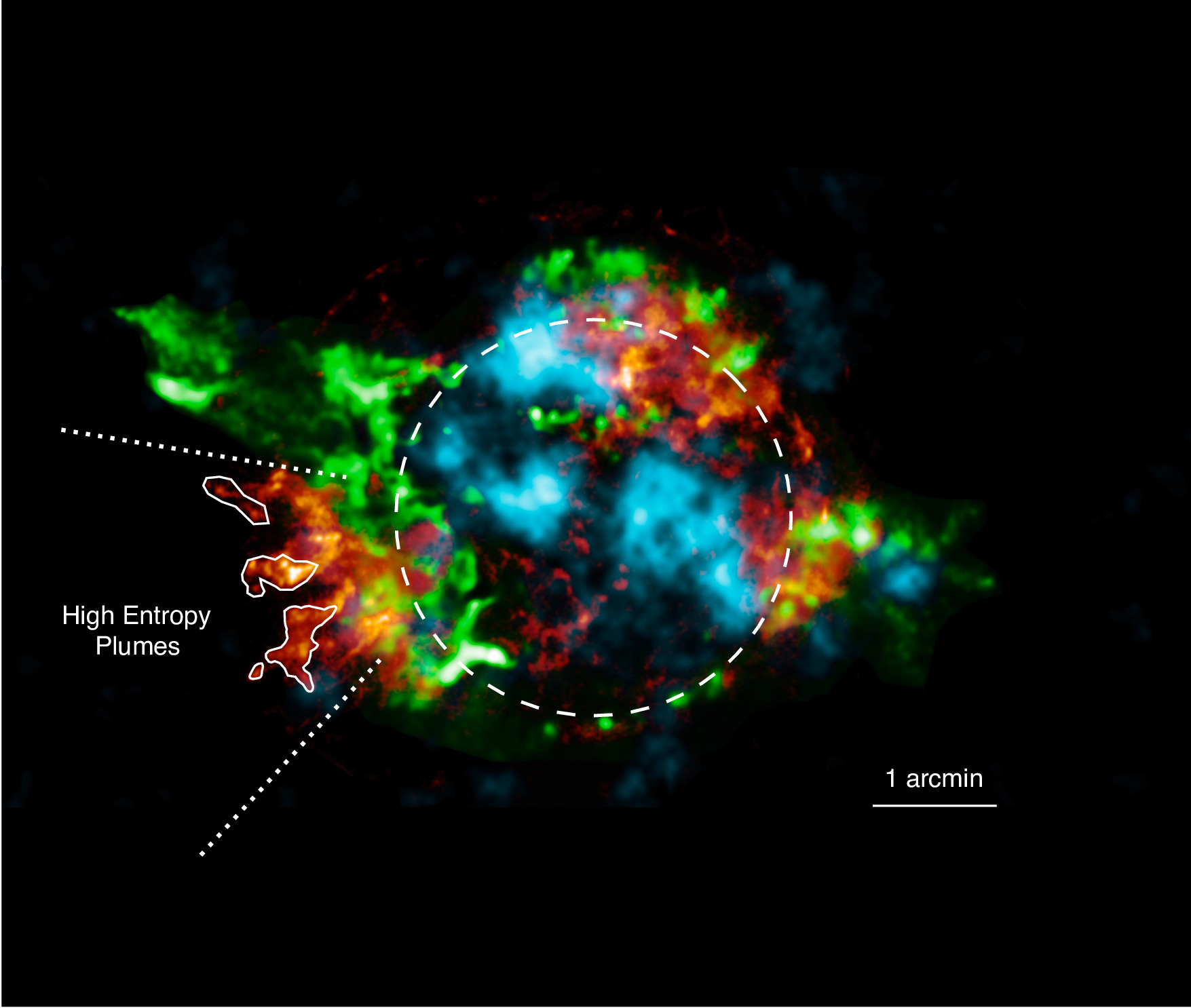}}
\caption{
\textbf{Asymmetric distribution of elements in Cassiopeia A supernova remnant.}
The iron ejecta (red) are popping out in the southeastern direction. The {\it Chandra} ratio map of the Si/Mg band is shown in green color where the jet structures can be seen at the northeastern and southwestern directions. The $^{44}$Ti observed by {\it NuSTAR}\cite{2014Natur.506..339G,2017ApJ...834...19G} is shown in blue, which is concentrated in the central region. The broken circle shows the mean location of the reverse shock\cite{2001ApJ...552L..39G}. From the white contour region, the X-ray spectrum in this study was extracted. At the southeastern iron-rich region, the firm detection of $^{44}$Ti has not been reported\cite{2017ApJ...834...19G}.}
\label{fig:f1}
\end{figure}

The young supernova remnant Cassiopeia A ($\sim$350 yr) exhibits asymmetric and bright Fe-K distribution\cite{2000ApJ...528L.109H} making it a unique target in our galaxy to examine details of the core-collapse explosion mechanism. In particular, the remnant shows protruding iron-rich fingers at the southeastern region\cite{2003ApJ...597..362H} (Fig.~\ref{fig:f1}). A key premise, derived from these results, is to relate the iron-rich ejecta to strong asymmetric phenomona in the central core of the explosion. 

Detection of some ``$\alpha$-rich freeze-out" products in the iron-rich ejecta would allow a direct connection with the predicted high-entropy plumes. $\alpha$-rich freeze out is a nuclear burning regime that controls explosive nucleosynthesis at high-entropy (especially for CC SNe). In the high-entropy nuclear burning regime, abundant $\alpha$ particles ($^{4}$He) are captured by heavy elements. Thus, the coexistence of the main burning product, iron ($^{56}$Fe after decay of $^{56}$Ni), and other $\alpha$ elements (e.g., $^{44}$Ti, $^{64}$Zn) in the same physical location would be strong evidence for the high-entropy process\cite{1998ApJ...492L..45N}. The $^{44}$Ti decay line has been detected in Cassiopeia A\cite{1994A&A...284L...1I,2014Natur.506..339G,2017ApJ...834...19G}, but the weak correlation in the spatial distribution between Fe and Ti has complicated the picture\cite{2014Natur.506..339G} (Fig.~\ref{fig:f1}). 

In addition to the most famous $\alpha$-rich freeze-out product, radioactive $^{44}$Ti, other rare stable elements synthesized through subsequent captures of $\alpha$-particles, such as stable Ti, Cr and Zn (see Methods for the detailed isotopes), can also be used to characterize the high-entropy nuclear burning regime\cite{2001ApJ...555..880N,2018ApJ...852...40W}. X-ray thermal emission lines from these and other elements provide direct measurement of the relative element compositions within a specific parcel of shocked plasma. This is indeed impossible with the $\gamma$-ray lines from $^{44}$Ti, since there are no similar detectable lines arising from other species via the same emission mechanism.  In this letter, we report the discovery of $\alpha$-rich freeze-out products, namely stable titanium and chromium, from the iron-rich ejecta of Cassiopeia A.

\begin{figure}
\internallinenumbers % required by nature
\begin{center}
\includegraphics[width=12cm, bb=0 0 1123 1380]{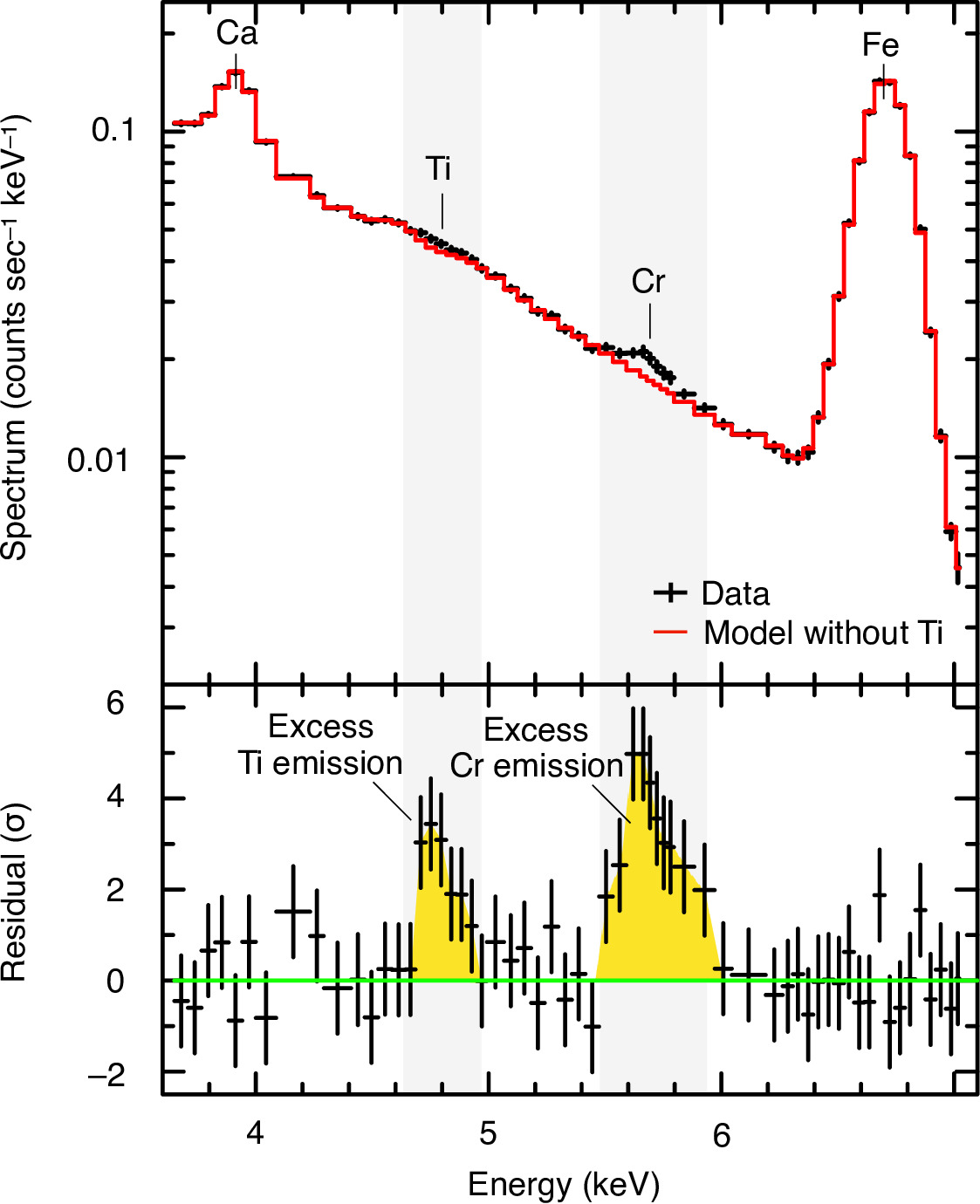}
\end{center}
\caption{
\textbf{The X-ray spectrum of the southeastern iron-rich ejecta.}
All of the {\it Chandra} data observed in 2000--2018 are combined and shown as black data points with 1$\sigma$ error bars. The modelled spectrum (red solid line) was obtained using an ionizing plasma model (vvpshock in Xspec) and the atomic databace AtomDB (Method). In $\sim$4.7--4.8 keV and $\sim$5.5--5.9 keV (gray and yellow shadow), the X-ray line features that we analyzed to come from the shocked titanium and chromium can be identified. Note that titanium and chromium are not included in the plasma model used in this figure, in order to clearly demonstrate a need for these elements to explain the observed spectrum. Once titanium and chromium are included in the plasma model, no significant residual is left (Methods).}
\label{fig:f2}
\end{figure}

\begin{figure}
\internallinenumbers % required by nature
\resizebox{\hsize}{!}{\includegraphics[bb=0 0 1288 937]{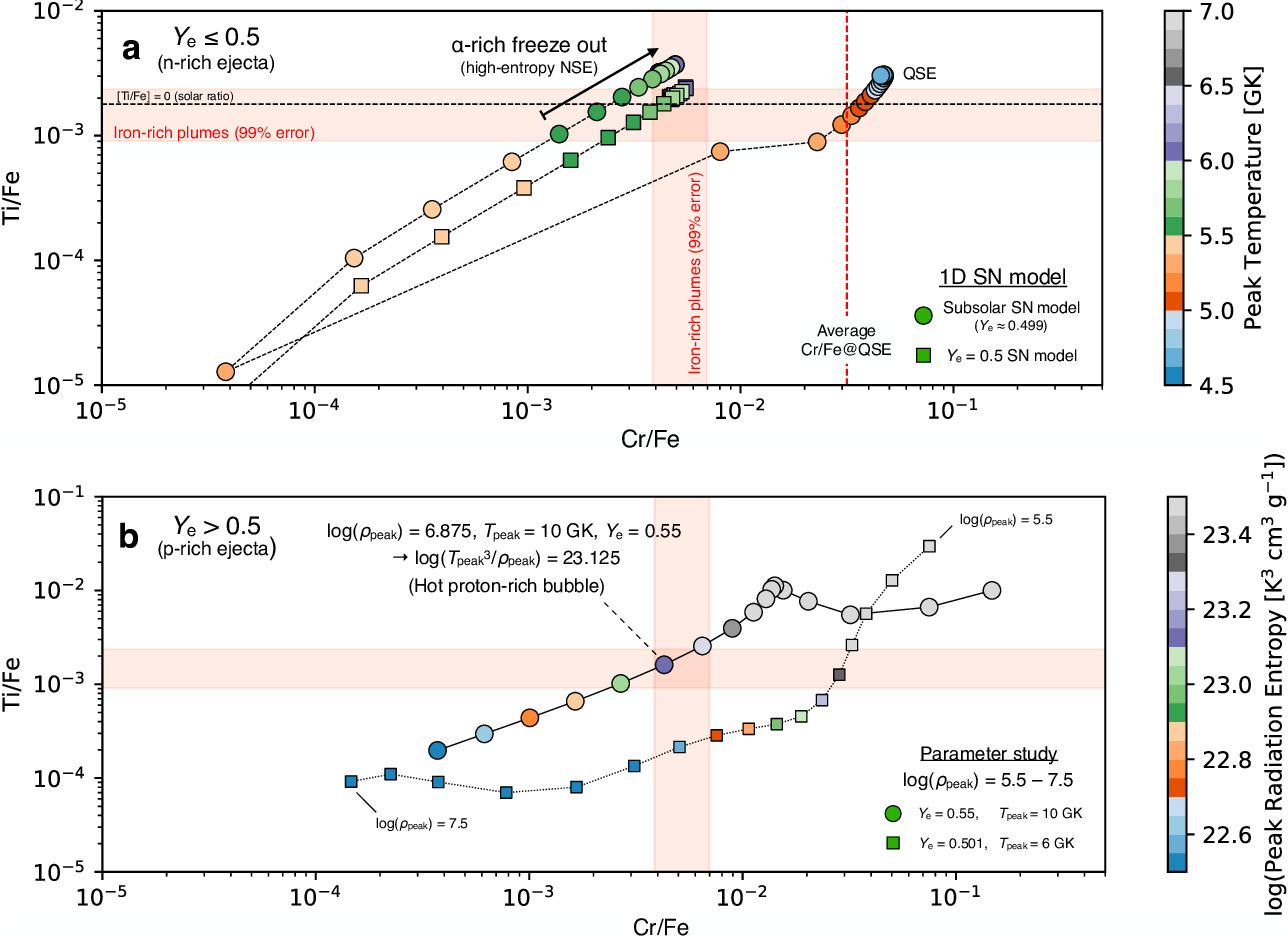}}
\caption{
\textbf{Comparisons of the observed mass ratios with those by theoretical calculations.}
{\bf a.} The Ti/Fe v.s. Cr/Fe mass ratios. The horizontal broken lines show [Ti/Fe] = 0 (the solar ratio\cite{1989GeCoA..53..197A}). The faint orange areas are the observed mass ratios (99\% confidence level, $\Delta\chi^2$ = 6.64). The colored points show the mass ratios in our nucleosynthesis calculations for core-collapse supernovae. We assumed a 15 $M_{\odot}$ progenitor with the sub-solar metallicity\cite{2020ApJ...893...49S} of $Z = 0.5 Z_{\odot}$ and a high explosion energy of 3$\times$10$^{51}$ erg for the circle data. In the box data, the lepton fraction at the $\alpha$-rich freeze out is modified to $Y_{\rm e} = 0.5$.
{\bf b.} The circle data show a parameter study of nucleosynthetic outputs (Ti/Fe and Cr/Fe) for hot ($T_{\rm peak} =$ 10 GK) and proton-rich ($Y_{\rm e}$ = 0.55) environment while changing the peak density from 10$^{5.5}$ g cm$^{-3}$ to 10$^{7.5}$ g cm$^{-3}$. In the box data, the nucleosynthesis calculations with a slightly proton-rich environment ($Y_{\rm e}$ = 0.501) are shown.
}
\label{fig:f3}
\end{figure}

We find large residuals from the reference model spectrum around 4.7--4.8 keV and 5.6--5.8 keV in the southeastern iron-rich ejecta region using {\it Chandra} (Fig.~\ref{fig:f2}), which can be explained only by K$\alpha$ emission from He-like Ti and Cr ions. Using an ionizing plasma model that includes these lines, we find that the spectrum is well fitted while the residual features are eliminated (see Methods). We compute observed mass ratios of Ti/Fe = 0.09--0.24\% and Cr/Fe = 0.39--0.70\% (these are 99\% confidence level ranges) in this region. These mass ratios require the yields of Fe-peak elements produced by $\alpha$-rich freeze out process (Fig.~\ref{fig:f3}a, Method). There exists some contamination from unrelated Si-rich ejecta within the selected region, but it is negligible in deriving the Ti and Cr abundances (Methods). While stable Ti and Cr are produced at both the $\alpha$-rich freeze-out and lower-temperature quasi-equilibrium (QSE) layers, the low Cr/Fe ratio rules out an origin in the QSE layer (Fig.~\ref{fig:f3}a). Thus, we conclude that the iron-rich fingers were produced at the high-entropy nuclear burning region. 

The mass ratios among the $\alpha$-rich freeze-out elements are specified by only a few parameters: the lepton (electron) fraction (the average electron number per baryon), the peak density, the peak temperature, the rate at which the temperature/density change and the freezeout timescale\cite{1973ApJS...26..231W,2020ApJ...895...82V}. This means, in principle, that the observed mass ratios can provide estimates of key physical parameters (i.e., the peak radiation entropy, $s_{\rm peak} \equiv T_{\rm peak}^{3}/\rho_{\rm peak}$ and the lepton fraction, $Y_{\rm e}$) of the convective supernova engine. In fact, some recent simulations predict that strongly neutrino-heated plumes are high-entropy and proton-rich\cite{2006A&A...447.1049B,2018ApJ...852...40W,2020MNRAS.491.2715B}. 

In Fig.~\ref{fig:f3}b, we show the nucleosynthetic outputs for an extremely hot ($T_{\rm peak} =$ 10 GK) burning zone with $Y_{\rm e} >$ 0.5 that can be achieved only in such neutrino-heated plumes (Methods). We find that the high-entropy (log($T_{\rm peak}^{3}/\rho_{\rm peak}$ [K$^3$ cm$^3$ g$^{-1}$]) $\gtrsim$ 23) and proton-rich ($Y_{\rm e} =$ 0.55) environment can reproduce the observed mass ratios very well.  These physical parameters are very similar to those of hot plumes in multi-dimensional simulations\cite{2018ApJ...852...40W,2020MNRAS.491.2715B,2020ApJ...895...82V}. In another possible scenario, we can also explain the mass ratios with the material immediately above the hot plumes that is characterized by lower entropy and higher density ($T_{\rm peak} \approx$ 6 GK, $\rho_{\rm peak} \approx$ 10$^7$ g cm$^{-3}$, log($T_{\rm peak}^{3}/\rho_{\rm peak}$) $\approx$ 22.3) than the hot plumes themselves. In this case, the lepton fraction needs to be modified to be $Y_{\rm e} \approx$ 0.5 (see curves with square points in Fig.~\ref{fig:f3}a). Such conditions are found in recent neutrino-driven SN simulations\cite{2020MNRAS.491.2715B} where the neutrino exposure is key to changing the lepton fraction from the original stellar value ($Y_{\rm e} <$ 0.5). Both scenarios are the direct consequence of the neutrino-driven convective engine. Distinguishing the two scenarios further is not straightforward, but the first scenario, i.e., the nucleosynthesis products within the hot plumes, is preferred; a drawback of the second scenario is that the Mn/Fe ratio does not support the $Y_{\rm e}$ modification, while it does agree well with the proton-rich case (Methods).  Other elements such as Ni that are sensitive to the $Y_{\rm e}$ value will become useful in the near future with improved X-ray detector technology (Methods).

Based on Doppler velocity and proper motion measurements, the total space velocity of the iron-rich ejecta is estimated to be $\gtrsim$ 4,000 km s$^{-1}$. In multi-dimensional simulations\cite{2003A&A...408..621K,2010ApJ...714.1371H,2017ApJ...842...13W}, iron-rich clumps produced by the combination of convective overturn and the growth of Rayleigh-Taylor instabilities during the explosion have maximum velocities of $\approx$4,000--5,000 km s$^{-1}$. Thus, the observed kinematics of the high-entropy iron-rich ejecta agree well with theoretical predictions.

The high-entropy iron-rich plumes appear to smoothly connect with the characteristic 3D structures (i.e., bubble-like interior and outer ring-like structures) of the ejecta in Cassiopeia A\cite{2013ApJ...772..134M,2015Sci...347..526M}. Recent multi-dimensional simulations of the remnant formation based on a neutrino-driven SN explosion model\cite{2017ApJ...842...13W,2021A&A...645A..66O} demonstrated that these observational characteristics of the remnant can be naturally explained by convective overturning in the neutrino-heating layer and the standing accretion shock instability. Especially, the simulated large cavities along the direction of propagation of the iron-rich plumes bear a remarkable resemblance to the structure around the southeastern iron-rich region in Cassiopeia A. Thus our results on the iron-rich plumes and their proposed formation process agree well with existing evidence from multi-wave band data of the global structure of the remnant interior.

While the detection of $^{44}$Ti (a representative isotope produced in $\alpha$-rich freeze out) has been reported, the poor spatial correlation between the  $^{44}$Ti and iron-rich X-ray ejecta has been puzzling\cite{2014Natur.506..339G} (Fig.~\ref{fig:f1}).  A possible explanation is that most of the iron processed by the $\alpha$-rich freeze out has not yet been heated to X-ray-emitting temperatures by the reverse shock. In addition, {\it NuSTAR}'s sensitivity to blue-shifted $^{44}$Ti lines is worse due to an instrumental feature in the mirror reflectivity (Pt 78.4 keV K-edge), hence only upper limits\cite{2017ApJ...834...19G}  could be set on the $^{44}$Ti emission in the blue-shifted iron-rich regions\cite{2002A&A...381.1039W,2010ApJ...725.2038D}  (see Method). Our results show not only that iron and the $\alpha$-rich freeze out elements (Ti and Cr) indeed spatially coexist in the plumes, but also that this region lies at the outermost edge of the remnant in the southeast. Interestingly, the proton-rich ejecta we propose for the iron-rich plumes typically contains less $^{44}$Ti than in neutron-rich ejecta\cite{2018ApJ...852...40W} and indeed our estimates of the  $^{44}$Ti mass in the Fe-rich plumes falls below the {\it NuSTAR} upper limits (see Methods). The finding here supports the basic picture of the convective SN explosion mechanism, with a need of strong asymmetry, therefore shedding light on an important decades-long unresolved problem in astrophysics.

Recent {\em Hitomi} SXS observations have demonstrated that high-resolution X-ray spectroscopy with X-ray calorimeters is a powerful tool for measuring the  abundances of rare elemental species\cite{2017Natur.551..478H}. Upcoming X-ray calorimeter missions (e.g., {\it XRISM}\cite{2020SPIE11444E..22T}, {\it Athena}\cite{2018SPIE10699E..1GB}) will provide us a great opportunity to investigate the role of the high-entropy nuclear burning in the Universe using stable titanium (and also zinc) as a new tool.

\newpage
%%%%%%%%%%%%%%%%%%%%%%%%%%%%%%%%%%%%%%%%%%%%%%%%%%%%%%%%%%%%%%%%%%%%%%%%%%%%%%%

%%%%%%%%%%%%%%%%%%%%%%%%%%%%%%%%%%%%%%%%%%%%%%%%%%%%%%%%%%%%%%%%%%%%%%%%%%%%%%%

%%%%%%%%%%%%%%%%%%%%%%%%%%%%%%%%%%%%%%%%%%%%%%%%%%%%%%%%%%%%%%%%%%%%%%%%%%%%%%%

%%%%%%%%%%%%%%%%%%%%%%%%%%%%%%%%%%%%%%%%%%%%%%%%%%%%%%%%%%%%%%%%%%%%%%%%%%%%%%%

\clearpage

\noindent{\Large\bf Methods} \vspace{0.4cm}\\
%%%%%%%%%%%%%%%%%%%%%%%%%%%%%%%%%%%%%%%%%%%%%%%%%%%%%%%%%%%%%%%%%%%%%%%%%%%%%%%
\textbf{Previous studies of the ejecta distribution in Cassiopeia A.}
Cassiopeia A, as a young, nearby and bright remnant of a core-collapse supernova, is a unique object to probe the explosion mechanism of massive stars. In particular, the asymmetric ejecta distribution and kinematics in the remnant that have been supported from various observations provide an important clue to understand the asymmetry in the explosion itself. Here we summarize the characteristic ejecta distribution in the remnant investigated in previous studies in order to clarify the novelty of this research. 

The high-velocity optical knots have clearly shown the asymmetric explosion in Cassiopeia A\cite{2001ApJS..133..161F,2006ApJ...645..283F,2016ApJ...818...17F,2001AJ....122..297T}. In particular, the S-rich optical knots in the NE jet and SW counterjet regions have high velocities of $>$ 10,000 km s$^{-1}$, which is a unique feature connected to the explosion mechanism. These jet-like structures are kinematically and chemically distinct from the rest of the remnant. Motivated by this structure, there have been some previous papers\cite{2008ApJ...677.1091W} proposing a jet-induced explosion scenario to explain the NE/SW jet-like structures and the southeastern Fe-rich ejecta. However, the distribution of $^{44}$Ti\cite{1994A&A...284L...1I,2001ApJ...560L..79V,2006ApJ...647L..41R,2014Natur.506..339G,2017ApJ...834...19G} does not support highly asymmetric bipolar explosions resulting from a fast-rotating progenitor\cite{2014Natur.506..339G},  even though it infers (mildly) asymmetric explosion in Cassiopeia A indirectly.

The neutrino-driven convection is a key mechanism to explode stars by neutrino-driven explosion. Especially, a recent simulation succeeded to reproduce the observed SN properties well, even the explosion energy\cite{2020arXiv201010506B}, thus observational verification of this mechanism would be timely and significant. To probe this, the Fe-rich ejecta could be a unique target. The formation process of the southeast Fe-rich ejecta\cite{2000ApJ...528L.109H,2000ApJ...537L.119H,2002A&A...381.1039W,2003ApJ...597..362H,2004ApJ...615L.117H} has been debated until this research. Hughes et al. (2000)\cite{2000ApJ...528L.109H}, for the first time, has proposed that the Fe-rich ejecta could be originated from the rising bubbles in the neutrino-driven convection layer\cite{1994ApJ...435..339H,1995ApJ...450..830B,2007PhR...442...38J,2012PTEP.2012aA309J,2021Natur.589...29B} during the supernova explosion. Also, Hwang et al. (2003)\cite{2003ApJ...597..362H} proposed a possibility that the Fe-rich ejecta were produced by $\alpha$-rich freezeout, through the same nucleosynthesis process with the formation of the Fe-rich region, based on the physical properties investigated by X-rays. However, the mismatch of Ti-rich and Fe-rich ejecta regions could not support the $\alpha$-rich freezeout origin well\cite{2014Natur.506..339G,2017ApJ...834...19G}. Our results solve this apparent contradiction, and observationally verify the existence of the neutrino-driven convection based on the discovery of the high-entropy plumes.

The origin of the Fe-rich ejecta should be closely connected to its kinematic structure. DeLaney et al. (2010)\cite{2010ApJ...725.2038D} argued that the Fe-rich ejecta occupies a hole in the Si-group emission and does not represent ``overturning'', as previously thought\cite{2000ApJ...528L.109H}. Instead of overturning, the authors proposed interaction with the circumstellar medium to reproduce the protruding Fe-rich ejecta. If this is the case and there is no inversion of the ejecta layers, the outermost tip of the Fe-rich ejecta should be produced by QSE (and we would see an O-burning layer on the outside of it). It would be not natural that all the outer layers (e.g. QSE, O-burning and so on) around the Fe-rich ejecta are hided only by the cross-section effect. This is in tension with the nucleosyntgetic origin of the Fe-rich ejecta we constrained in this work. However, the structure of the Fe-rich ejecta can indeed be well reproduced also by multi-dimensional simulations\cite{2003ApJ...598.1163M,2016ApJ...822...22O,2021A&A...645A..66O} even without the CSM interaction. Especially, a recent SNR simulation\cite{2021A&A...645A..66O} based on a neutrino-driven explosion have succeeded to reproduce naturally the characteristic ejecta structures in Cassiopeia A without any special CSM interaction. The picture here is consistent with our conclusion that the outermost tip of the southeast Fe-rich ejecta was produced by $\alpha$-rich freezeout. In summary, we conclude that the formation scenario of the Fe-rich ejecta with the inversion of ejecta layers during the SN explosion for Cassiopeia A is supported by all the available observational data. 

The bubble-like interior observed in Cassiopeia A\cite{2015Sci...347..526M} also provides strong evidence of turbulent mixing processes during the explosion. At optical and infrared wavelengths, arc- and ring-like structures of shocked ejecta have been observed\cite{1995ApJ...440..706R,1995AJ....109.2635L,2014MNRAS.441.2996A,2010ApJ...725.2038D}. The bubble-like morphology revealed by near-infrared observations of unshocked ejecta smoothly connects to these structures. This internal structure indirectly supports the existence of a ``Ni-bubble'' effect during the explosion\cite{1989ApJ...341L..63A,1993ApJ...419..824L,2001ApJ...557..782B,2013ApJ...773..161O}.
 The Ni-bubble effect has been used to explain observations of SN 1987A indicating extensive mixing of chemical species in the envelope of the progenitor star during the explosion. Early simulations of the effect were started by artificially seeding Rayleigh-Taylor instabilities (RTIs) in the mantle and envelope of the progenitor and following their evolution until shock breakout from the stellar surface\cite{1989ApJ...341L..63A,1991ApJ...367..619F,1990ApJ...358L..57H,1991A&A...251..505M,1991ApJ...370L..81H,1993ApJ...419..824L}. More recent simulations consistently connect the seed asymmetries arising from convective flow in the neutrino-heated bubble and by the SASI during the explosion\cite{2003A&A...408..621K,2007PhR...442...38J,2010ApJ...714.1371H,2012PTEP.2012aA309J,2021MNRAS.tmp..147G}. This is the same scenario as explaining the formation of the Fe-rich ejecta\cite{2000ApJ...528L.109H}. Thus, the origin of the bubble-like interior in the remnant could smoothly connect with the formation process of the protruding Fe-rich ejecta and the recent simulation study supports this picture\cite{2021A&A...645A..66O}.

As summarized above, the previous studies focused on the ejecta distribution in Cassiopeia A to tackle to the explosion mechanism. In the present work, we introduce a new dimension, namely the nucleosynthetic perspective of the high-entropy bubbles as a key ingredient of the neutrino-driven explosion. We for the first time demonstrated that the X-ray observation of the Fe-rich ejecta can provide us the mass fractions of the key elements, which allow us to probe local physical parameters (i.e. lepton fraction, peak temperature, peak density and so on) around the convective SN engine. Currently, this is the only way to measure these parameters in a SN explosion. 

\noindent\textbf{Chandra observations and data reduction.} Cassiopeia A has been observed with {\it Chandra} ACIS-S several times since the launch\cite{2000ApJ...528L.109H,2011ApJ...729L..28P,2014ApJ...789..138P,2017ApJ...836..225S,2018ApJ...853...46S}. We used all ACIS-S observations from 2000 to 2018, with a total exposure of $\sim$1.57 Ms. We reprocessed the event files (from level 1 to level 2) to remove pixel randomization and to correct for CCD charge transfer efficiencies using CIAO\cite{2006SPIE.6270E..1VF} version 4.6 and CalDB 4.6.3. The bad grades were filtered out, and good time intervals were reserved.

Extended Data Fig.~\ref{fig:E1}a shows a Fe-K/Si-K ratio map where the white contour is the region we extracted the spectrum. In order to extract a Fe-rich spectrum, only bright regions were selected using this image. The position of the iron-rich ejecta we focused is shifting from year to year due to the bulk expansion of the material. Therefore, we carefully chose the regions for each epoch considering the shift to track the same material. The proper motions of the iron-rich structures were estimated to be $\sim$0.2$^{\prime\prime}$ yr$^{-1}$. With a similar shift to that measurement, we were able to correct the position of the regions well. As shown in Extended Data Fig.~\ref{fig:E1}b, we could track well the entire Fe-rich structure from epoch to epoch although the shape of the structure is slightly changing.

The shape of small structures at off-axis angle is distorted by the aberrations of Wolter type I optics. This may increase uncertainty for the region selection from epoch to epoch. On the other hand, the encircled energy radii for circles enclosing 90\% of the power at 6.4 keV around the Fe-rich region (the off-axis angle of $\sim$3 arcmin) is $\sim$2 arcsec\footnote{see Figure 4.12 \& 4.13 in https://cxc.harvard.edu/proposer/POG/html/chap4.html}, which is much smaller than the regions we used (see Extended Data Fig.~\ref{fig:E1}a). Therefore, the difference in the image distortion due to the off-axis effect from epoch to epoch can be ignored. In fact, a comparison between single epoch and multiple epoch measurements shows that the results are consistent, indicating that the region selection itself does not have a significant impact on the Ti measurements (see following sections and Extended Data Fig.~\ref{fig:E1}d,e).

We also analyzed the X-ray spectrum taken from a smaller region shown in Extended Data Fig.~\ref{fig:E1}a (white broken contour). We found that the change of the region boundaries does not change the Ti/Fe ratio (Extended Data Fig.~\ref{fig:E1}e). On the other hand, the Ca/Fe mass ratio decreases from 1.50--1.54\% to 1.07--1.20\% when we use the smaller region. The decrease of Ca/Fe is interpreted as a decrease of the contamination from the Si-rich component (see also ``The origin of the lighter elements in the iron-rich ejecta''). The unchanged Ti/Fe ratio implies that the Si-rich component is not the main component providing Ti. From these, we are confident that the region selection does not change our conclusion significantly.

\noindent\textbf{NuSTAR observations and data reduction.} In Figure \ref{fig:f1}, the same $^{44}$Ti image as in Grefenstette et al. (2017)\cite{2017ApJ...834...19G} was used. Cassiopeia A was observed with {\it NuSTAR} during the first 18 months of the {\it NuSTAR}mission with a total exposure time of 2.4 Ms. We reduced the {\it NuSTAR} data with the {\it NuSTAR} Data Analysis Software (NuSTARDAS) version 1.4.1 and {\it NuSTAR} calibration database (CALDB) version 20150316 to produce images, exposure maps, and response files for each telescope.

\noindent\textbf{Modeling of X-ray spectrum.} We extract the X-ray spectrum from the southeastern iron-rich structure (Fig.~\ref{fig:f2}). To model it, we use an absorbed thermal plasma model with a gsmooth model that is used for modeling the lines broadened by thermal broadening and/or multiple velocity components (= phabs$\times$gsmooth$\times$vvpshock in Xspec using AtomDB version 3.0.9). We assumed that the plasma parameters (e.g., ionization state, temperature, redshift) for each element are identical. Although the column density $n_{\rm H}$ is not sensitive to this energy band (3.7--7.1 keV), we fixed it to the typical value around this region\cite{2003ApJ...597..362H}. The best-fit parameters are summarized in Extended Data Fig.~\ref{fig:E1}d.

The residuals around $\sim$4.7--4.8 keV in Fig.~\ref{fig:f2} are very well explained by the Ti emissions of the thermal model (Extended Data Fig.~\ref{fig:E1}). The main difference in the models shown in these two figures is whether titanium is included (Extended Data Fig.~\ref{fig:E1}) or not (Fig.~\ref{fig:f2}). The significance of the Ti detection in the thermal fitting is $\approx$ 5.6$\sigma$ ($\Delta\chi^2 =$ 33.044). Even if we used only the data in 2004 (the net exposure time $\approx$ 980 ks), the significance level is still above 5$\sigma$ ($\Delta\chi^2 =$ 25.422). Even if we used an ionizing plasma model with a single ionization state (NEI model), the significance level of the line detection is almost the same ($\Delta\chi^2 =$ 25.219). We also analyzed all the data except 2004 (the net exposure time $\approx$ 590 ks) to check the consistency and again detected the Ti emissions at a 2.7$\sigma$ confidence level. All the Ti/Fe ratios in the different data sets are consistent with each other (see Extended Data Fig.~\ref{fig:E1}d,e), which indicates the robustness of our Ti measurements.

We also performed a simple Gaussian fitting for the residual line. Here we used a zgauss model in Xspec to express the blue-shifted line. Assuming the same blue-shift velocity as in the vvpshock model ($\sim -$2,860 km s$^{-1}$), the centroid energy of the residual line is estimated to be 4.74$\pm$0.03 keV (90\% confidence level, $\Delta\chi^2 = 2.706$) in the rest frame, which is consistent with the K$\alpha$ emissions from the He-like Ti ion (Extended Data Fig.~\ref{fig:E2}a). We note that there is almost no bright X-ray line around this energy band in the AtomDB database, except for the titanium. For the Gaussian fitting, the significance of the line detection is almost same as that in the thermal fitting ($\Delta\chi^2 = 26.997$).

We note that the energy of the He-alpha lines of Sc are 4.295 kev (resonance line) and 4.316 keV (intercombination line). This is close to the H-like Ca K$\alpha$ line energy (4.1 keV), but it would not be a source of confusion for the detection of stable Ti. On the other hand, we would note that the uncertainty on the line emissivities (i.e. uncertainty of atomic data base) could change the estimations of mass ratios.

The narrow fitting range from 3.7 to 7.1 keV does not affect our conclusion. Even if we expanded the fitting range, we could obtain good spectral fits (reduced $\chi^2 <$ 2) and the Ti/Fe mass ratios are consistent with each other (Extended Data Fig.~\ref{fig:E1}e). In Extended Data Fig.~\ref{fig:E2} bottom, we show the entire X-ray spectra fitted with different fitting range. In all the fittings, the entire spectra are roughly explained by the best-fit models. Here, elements in areas outside of the fitting range are fitted by eye and fixed. In Extended Data Fig.~\ref{fig:E2}d, we fit the spectrum up to 9.5 keV where the Ni emissions are included. The current modeling around 8.0--8.5 keV is not accurate enough (please also see ``Another strong evidence of $\alpha$-rich freezeout''). This is due to uncertainties of emissivities of Fe K$\beta$,$\gamma$,$\delta$,... emissions. To model a lack of these emissions, we added a Gaussian model in this fitting. In Extended Data Fig.~\ref{fig:E2}e, we fit the spectrum from 2.2 to 9.5 keV where the emissions from S to Ni are included (a complete spectral modeling for the entire energy band from 0.7 to 9.5 keV is too complicated to realize for now). Even in the wide range fitting, we could obtain a good fit (reduced $\chi^2 =$ 1.71). The best-fit parameters are summarized in Extended Data Fig.~\ref{fig:E2}f. In this case, we added a bremss model in Xspec because a low temperature component cannot be ignored\cite{2002A&A...381.1039W} due to expanding to the lower energy band. Considering the uncertainties of handling the multi-temperature modeling and also the uncertainties of modeling around the Ni emission line, we presented only the result for the 3.7-7.1 keV band in the main text.

\noindent\textbf{Estimation of the total iron/titanium mass in the iron-rich region.} We estimated that the three large iron-rich structures (white contour regions) at the southeastern region have a total volume of $\sim$0.141 arcmin$^{3}$ (= 0.6$^{\prime}$$\times$0.2$^{\prime}$$\times$0.2$^{\prime}$ for north one + 0.6$^{\prime}$$\times$0.3$^{\prime}$$\times$0.3$^{\prime}$ for middle one + 0.7$^{\prime}$$\times$0.3$^{\prime}$$\times$0.3$^{\prime}$ for south one), which corresponds to 4$\times$10$^{54}$ cm$^{-3}$ at the distance of 3.4 kpc. The emission measure, $n_{e} n_{\rm H} V$ is estimated to be 1.1$\times$10$^{57}$ cm$^{-3}$ from the fitting result (i.e., from the norm parameter in Extended Data Fig.~\ref{fig:E1}d). Assuming $n_{e} \approx 1.2 n_{\rm H}$ (for fully ionized H-dominant plasma with 10\% He) and [Fe/H]/[Fe/H]$_{\odot}$ = 8.7, the iron mass in the structures is estimated to be $\approx$10$^{-3}$ $M_{\odot}$. On the other hand, the total iron mass in the southeastern region is estimated to be 0.018 $M_\odot$ in previous studies\cite{2017ApJ...834...19G}, which is almost one order of magnitude higher than our estimation (probably it's due to some different assumptions). Compared to the total mass of all the synthesized iron in this remnant\cite{2012ApJ...746..130H}, the iron mass at the edge of the southeastern structures is not so large ($\sim$1--10\% of the entire iron ejecta). Here, the total stable titanium mass in the iron-rich region could be estimated to be 10$^{-6}$--10$^{-5}$ $M_{\odot}$ assuming Ti/Fe = 10$^{-3}$.

Extended Data Fig.~\ref{fig:E3} shows the $^{44}$Ti upper-limit map by {\it NuSTAR}\cite{2017ApJ...834...19G}. Around the southeastern iron-rich regions, the upper limits show 7$\times$10$^{-7}$ ph cm$^{-2}$ s$^{-1}$ for each box region (box id: 31, 39, 47), which corresponds to $^{44}$Ti mass of $<$ 5.8$\times$10$^{-6}$ $M_\odot$. Assuming the iron mass of 10$^{-3}$--10$^{-2}$ $M_{\odot}$ and the mass ratio of Ti/Fe = 10$^{-3}$ and $^{44}$Ti/Ti = 0.2 (Extended Data Fig.~\ref{fig:E4}), the $^{44}$Ti mass in the iron-rich region is estimated to be 2$\times$10$^{-7}$--2$\times$10$^{-6}$ $M_{\odot}$, which is almost comparable or an order of magnitude lower than the {\it NuSTAR}'s upper limit. In the case of proton-rich ejecta, the synthesized amount of $^{44}$Ti is almost one order of magnitude less than this estimation\cite{2018ApJ...852...40W}. Therefore, especially in the proton-rich case, it would be reasonable that the {\it NuSTAR}'s observations could not detect the $^{44}$Ti line significantly from this region. In addition, recent multi-dimensional simulations\cite{2021A&A...645A..66O,2020ApJ...895...82V} showed that the mass ratio between $^{44}$Ti and $^{56}$Ni could vary by 3 orders of magnitude from region to region with an average mass ratio of $^{44}$Ti/$^{56}$Ni $\approx$ 10$^{-3}$. The regions with poor Ti production rates, such as those appearing in the multi-dimensional simulations, could also explain the undetected $^{44}$Ti in the iron-rich region.

\noindent\textbf{Nucleosynthesis model calculations.} We use nucleosynthesis models for core-collapse supernovae calculated in Sato et al. (2020)\cite{2020ApJ...893...49S}. Here we calculate the evolution of a 15 $M_\odot$ star with metallicity of $Z$ = 0.007 (0.5 $Z_\odot$) from hydrogen burning until the central temperature reaches 10$^{9.9}$ K. The calculation is performed using a 1D stellar evolution code, HOngo Stellar Hydrodynamics Investigator (HOSHI) code\cite{2016MNRAS.456.1320T,2018ApJ...857..111T,2019ApJ...871..153T,2019ApJ...881...16Y}. Detailed parameter sets of the stellar models are the same as Set L in Yoshida et al. (2019)\cite{2019ApJ...881...16Y}. Nucleosynthesis of 300 isotopic species is also calculated within the stellar evolution code. The metallicity dependence of the mass-loss rate is $Z^{0.85}$ for main-sequence stars and $Z^{0.5}$ for yellow and red supergiants\cite{2013A&A...558A.103G}.

The simulation of the supernova explosion is performed with a PPM hydrodynamics code\cite{1984JCoPh..54..174C,2005ApJ...619..427U}, assuming a spherically symmetric explosion. The explosion energy $E_{\rm exp}$ is set to be 3$\times 10^{51}$ erg. The location of the mass cut is determined so that the ejected $^{56}$Ni mass is 0.07 $M_\odot$. After the supernova explosion simulations, the explosive nucleosynthesis calculations are performed in a postprocessing step.
Radioactive decays in the supernova ejecta after 350 yr are also taken into account.

As shown in Extended Data Fig.~\ref{fig:E4}, the region where Fe (though $^{56}$Ni) is abundantly produced is divided into two layers; the alpha-rich freezeout as characterized by the higher temperature (or entropy), producing mainly Fe-peak elements, and the incomplete QSE burning as characterized by the lower temperature, producing intermediate mass elements abundantly. In the one-dimensional model, the alpha-rich freezeout region is located in the deepest layer since higher entropy is realized in the inner layer in the one-dimensional model. However, this is not necessarily the case in realistic three-dimensional simulations\cite{2016ARNPS..66..341J,2020MNRAS.491.2715B}, since the penetration of the the high-entropy bubbles into the outer layers is essential to initiate the explosion.

Ti has 5 stable isotopes of which $^{48}$Ti is the most abundant comprising 74\% of
titanium atoms in terrestrial samples. In our SN model with $Y_{\rm e} \lesssim$ 0.5, $\sim$90\% of all Ti is $^{48}$Ti. The isotope of Ti most well
known to supernova researchers is $^{44}$Ti, which has a half life of 60
years, decaying to $^{44}$Sc by electron capture (and then to $^{44}$Ca by beta decay). In our models, all Ti isotopes are contained in ``Ti". The radioactive element $^{44}$Ti is also included, but stable Ti is dominant due to the decay for 350 yr. We summarized the stable isotopes of Ti, Cr and Ni for neutron-rich and proton-rich ejecta in Extended Data Fig.~\ref{fig:E5}b. The mass fractions among isotopes are different between them. In the case of Ti, the synthesized amount of $^{46,47,49}$Ti drastically increases in the proton-rich ejecta. The $Y_{\rm e}$ dependence of each isotope can be checked in Figure 10 of Wanajo et al. (2018)\cite{2018ApJ...852...40W}.

\noindent\textbf{Parameter studies of nucleosynthetic outputs in the peak temperature-density plane.} In the case of 1D SN calculations, it is limited to express an extreme environment in the hot bubbles produced by neutrino heating. The 1D SN calculations as described above deal only with the regions above the hot plumes, which are heated by the shock wave produced by the action of the hot plumes. To express the high entropy environment within the hot plumes seen in multi-dimensional models, another approach is necessary. In order to discuss the element composition in the hot bubbles, a parameter study beyond the parameter range in 1D SN models would be suitable. Magkotsios et al. (2010)\cite{2010ApJS..191...66M} have investigated the sensitivity of $^{44}$Ti and $^{56}$Ni synthesis for both dependencies over an extended parameter space. Here, we do a similar approach for investigating the yield of stable Ti and Cr, which would be useful for discussing the nucleosynthesis that cannot be expressed with the 1D SN models. The data in Fig.~\ref{fig:f3}b were made from nucleosynthesis calculations we introduce here.

Extended Data Fig.~\ref{fig:E5} shows the Ti/Fe and Cr/Fe mass ratios in the peak temperature-density plane. We can clearly see that both the mass ratios increase as the peak radiation entropy increases. This trend is the same as $^{44}$Ti\cite{2010ApJS..191...66M}, therefore these elements can be an alternative tool for probing the high-entropy process. 
In computing this, we need to specify the thermodynamic trajectory for the hot bubbles. Since the time scale for the expansion and cooling there (i.e., the rate of the change in the temperature and density) should be similar to the innermost zone seen in the 1D simulation, 
we first extrapolate the thermodynamic trajectory from the innermost zones of the 1D model to match the peak temperature to a given peak temperature for the hot plumes. Then, we multiply a constant value to the density trajectory to match to the density condition for the hot plumes. Radioactive decays in the supernova ejecta after 350 yr are also taken into account.

In Extended Data Fig.~\ref{fig:E6}, we show the Ti/Fe and Cr/Fe mass ratios that are calculated with some situations that are not considered in Extended Data Fig.~\ref{fig:E5}. Extended Data Fig.~\ref{fig:E6}a shows the mass ratios calculated with $Y_{\rm e} =$ 0.53--0.58 and $T_{\rm peak} =$ 10 GK. In multi-dimensional simulations, the lepton fraction in the innermost proton-rich ejecta whose progenitor mass is in 15--20 $M_\odot$ has a wide range from $Y_{\rm e} =$ 0.5 to 0.6\cite{2018ApJ...852...40W,2018MNRAS.477.3091V}. As shown in Fig.~\ref{fig:f3}b, slightly proton-rich ejecta could not reproduce the observed mass ratios. On the other hand, we found that the proton-rich ejecta with $Y_{\rm e} \gtrsim$ 0.51 can reproduce the observed mass ratios. When the $Y_{\rm e}$ increases to a certain extent, we can no longer use these mass ratios to distinguish the $Y_{\rm e}$ differences (Extended Data Fig.~\ref{fig:E6}a).

The yields of the $\alpha$-rich freezeout are very sensitive to the thermodynamic evolution of the material, especially for intermediate mass elements like Ti and Cr. Therefore, we also investigated nucleosynthetic outputs using different thermodynamic evolution. In Extended Data Fig.~\ref{fig:E6}b, we show the Ti/Fe and Cr/Fe mass ratios produced by power-law thermodynamic trajectories. Here we calculate the evolution of temperature and density using Eq.~(5) in Magkotsios et al. (2010)\cite{2010ApJS..191...66M}. As a result, we found that the yields of intermediate mass elements by the power-law thermodynamic trajectories need more high-entropy to reproduce the observation. Thus, we note that the current estimation of the radiation entropy (i.e. peak temperature and density) has a certain uncertainty. Special thermodynamic evolutions, such as those realized in multidimensional simulations\cite{2018ApJ...852...40W,2020ApJ...895...82V}, may further alter the synthesis of these intermediate mass elements. That will be our tasks in the future.

\noindent\textbf{The origin of the lighter elements in the iron-rich ejecta.} We found X-ray lines not only from the $\alpha$-rich freeze-out products, but also from the intermediate mass elements: IME (e.g., Si, S, Ar, Ca) in the spectrum of the iron-rich ejecta. In our analysis, the region we chose is relatively large to obtain enough photon statistics, which could produce contamination from some ejecta that are not iron-rich. Therefore, if the Ti and Cr we found were not the iron-rich ejecta origin, our conclusion could change. We here confirm that the effect of this contamination is not significant to our conclusion.

In Extended Data Fig.~\ref{fig:E8}a, we newly defined a Si-rich ejecta region to compare with the iron-rich ejecta. The Si-rich region is adjacent to the iron-rich ejecta. Extended Data Fig.~\ref{fig:E8}b shows the X-ray spectrum extracted from the Si-rich region, which is well explained by a thermal plasma model (phabs$\times$vvpshock). The best-fit parameters are summarized in Extended Data Fig.~\ref{fig:E8}e. Using the plasma model, we attempted to model the X-rays from the lighter elements in the iron-rich region (Extended Data Fig.~\ref{fig:E8}c). As a result, we found that the Si-rich region reproduces the IME lines up to Ca, while the lines from Fe-peak elements and continuum are dominanted by the Fe-rich region. Thus, it was confirmed that the lighter elements very likely originate in the contamination from faint Si-rich emissions along the same line of sight.

What is the nucleosynthetic origin for the Si-rich ejecta? The mass fractions among them will answer it. Extended Data Fig.~\ref{fig:E8}d shows a relation between Ca/Si and Fe/Si mass ratios. The observed Ca/Si and Fe/Si mass ratios in the Si-rich ejecta are well consistent with those at $T_{\rm peak} \approx$ 4.5 GK. Therefore, we conclude that the Si-rich ejecta have been reprocessed through the QSE (incomplete Si burning) layer (see Extended Data Fig.~\ref{fig:E4}). Around this peak temperature, a large amount of Ti and Fe can not be produced. This means that the contamination from the Si-rich ejecta to the Ti and Fe emissions in the iron-rich ejecta is negligible. Given that this contamination would not affect the fluxes of the Ti, Cr and Fe lines, the Ti/Fe and Cr/Fe are hardly affected. In fact, we confirmed that the derived ratios of Ti/Fe and Cr/Fe are nearly unchanged even if we subtracted the Si-rich model from the Fe-rich spectrum.

\noindent\textbf{Another strong evidence of $\alpha$-rich freeze out.} A strong line feature around 7.8 keV in the spectrum of the iron-rich structure (see Extended Data Fig.~\ref{fig:E2} bottom, Extended Data Fig.~\ref{fig:E8} and Extended Data Fig.~\ref{fig:E10}) implies a large abundance of Ni. The stable Ni (= $^{58}$Ni and $^{60}$Ni) is the second abundant element in the $\alpha$-rich freeze out (Extended Data Fig.~\ref{fig:E4}). The existence of Ni in the iron-rich ejecta strongly supports the $\alpha$-rich freeze-out origin.

Extended Data Fig.~\ref{fig:E9} shows a relation between the Ni/Fe and Cr/Fe mass ratios in the CC SN ejecta. We found that the best-fit Ni/Fe mass ratios estimated in the iron-rich ejecta is above a few percent, which can be archived only at high peak temperature above 5.5 GK. On the other hand, we note that there are large uncertainties of the current modelings\cite{2017Natur.551..478H}. In particular, both the Ni K$\alpha$ emissions and Fe K$\beta$ emissions contribute to the 7.8 keV line feature. The emissivities of these emissions vary from atomic code to atomic code, and thus the derived value of Ni/Fe has a large uncertainty (see blue and red broken lines in Extended Data Fig.~\ref{fig:E9}). Further investigations for Ni, taking these uncertainties into account, will be our future works. For example, high X-ray resolution spectroscopy will help us to measure the element abundances accurately even up to such rare metals. Extended Data Fig.~\ref{fig:E10} bottom shows a comparison of spectra between XRISM and {\it Chandra}. As in the figure, we can separate fine structures in the X-ray spectrum in the near future, which will provide us the robust element abundance measurements.

The accurate Ni measurement will help us to estimate the lepton fraction in the ejecta. In proton-rich side, $^{58,60}$Ni are the most abundant in stable nickel isotopes (Extended Data Fig.~\ref{fig:E5}b) where the Ni/Fe mass ratio is $>$ 10\%. In proton-rich side, $^{60}$Ni is the most abundant in stable nickel isotopes (Extended Data Fig.~\ref{fig:E5}b) where the Ni/Fe mass ratio is $\sim$3--5\% (see Extended Data Fig.~\ref{fig:E5}b and Extended Data Fig.~\ref{fig:E9}). In the current spectral fitting, the model in SPEX does a better job of reproducing the spectrum (at least, SPEX does not need the Gaussian line used in Xspec), but still we found some residuals of the fitting. Here, the best-fit Ni/Fe mass ratio with SPEX is $\approx$7\%, which is almost in between the proton-rich and neutron-rich ejecta (Extended Data Fig.~\ref{fig:E9}b). In addition, the Ni/Fe dependence on $Y_{\rm e}$ is not so sensitive in the proton-rich side (Extended Data Fig.~\ref{fig:E9}b). Therefore, a tight constrain of Ni/Fe beyond the current measurement will be needed to determine the $Y_{\rm e}$. At least, some updates of the atomic codes and the spectral resolution as done by {\it Hitomi}\cite{2018PASJ...70...12H} will be needed for such a measurement. And, we hope that {\it XRISM} will do this (Extended Data Fig.~\ref{fig:E10} bottom).

\noindent\textbf{The amount of Manganese in the Fe-rich ejecta.}
In contrast of the Cr production, the Mn is more effectively produced in the proton-rich side\cite{2018ApJ...852...40W}. This tendency would help us to discriminate between the proton-rich and neutron-rich ejecta. In Extended Data Fig.~\ref{fig:E7}, we found that the high-entropy proton-rich ejecta could reproduce the observed Mn/Fe ratio very well.
On the other hand, the Mn/Fe in the neutron-rich ejecta is out of the observed range. In addition, we needed a modified $Y_{\rm e}$ of 0.5 to reproduce both Ti/Fe and Cr/Fe as shown in the main text (see Fig.~\ref{fig:f3}). This modification (increase of $Y_{\rm e}$) suppresses the production of neutron-rich elements like Mn, which provides a further deviation from the observed ratio (see box data in Extended Data Fig.~\ref{fig:E7}). Thus, the high-entropy proton-rich ejecta are more favorable to explain the observation.

We note that the Mn production around the mass cut region is sensitive to $\nu$-process\cite{1990ApJ...356..272W,2008ApJ...672.1043Y}. Here, $^{55}$Co (= $^{55}$Mn) is produced through $^{56}{\rm Ni}(\nu,\nu^\prime p)^{55}{\rm Co}$ after the production of $^{56}$Ni through complete and incomplete Si burning. The contribution of $\nu$-process to the Mn production is significant especially in the neutron-rich side because the Mn is rarely synthesized by $\alpha$-rich freezeout with the neutron-rich environment ($Y_{\rm e} \lesssim$ 0.5). In the proton-rich side, the $\nu$-process causes almost no change in the amount of Mn. In $\nu$-process, the amount of Mn depends on the total neutrino energy in the supernovae. The total energy carried away by the neutrinos is determined by the binding energy released during the formation of a neutron star, where the binding energy ($E_{\rm binding}$) is described as $E_{\rm binding} \approx 1.5 \times 10^{53} ~(M/M_\odot)^2~{\rm ergs}$\cite{2001ApJ...550..426L}. In our calculations, we assumed a total neutrino energy of 3$\times$10$^{53}$ erg, which corresponds to the binding energy at a typical NS mass of $\sim$1.4 $M_\odot$. Even if we assumed a heavy NS with $\sim$2 $M_\odot$ ($E_{\rm binding} =$ 6$\times$10$^{53}$ erg), the Mn/Fe mass ratio produced by a SN model with $Y_{\rm e} =$ 0.5 is out of the 90\% error range of the observation. This still prefers the proton-rich environment for producing the Fe-rich ejecta.

\noindent\textbf{X-ray spectra in other Fe-rich regions.}
We have investigated X-ray spectra in three Fe-rich regions of Cassiopeia A to search for products by high-entropy burning, where the Fe-rich regions are located at the southeast, north and southwest regions (Extended Data Fig.~\ref{fig:E10}a). We found the spectrum in the north region has a significant Ti line with a confidence level of 3.3$\sigma$ (Extended Data Fig.~\ref{fig:E1}e), but we also found that it has a strong Cr line (Extended Data Fig.~\ref{fig:E10}b). The strong Cr line means that the majority of the ejecta were produced at the incomplete Si burning (QSE) layer (see Fig.~\ref{fig:f3} and Extended Data Fig.~\ref{fig:E4}). The QSE layer could also produce Ti, so it was difficult to discuss the Ti production at the high-entropy burning using this region. In the case of the southwest region, the non-thermal emissions are significant, which make it difficult to detect the Ti line. Thus, we discussed only the southeast Fe-rich region where we can purely discuss the high-entropy process in the main text.

\noindent\textbf{The Ti underproduction problem in the classical Galactic chemical evolution models.} Interestingly, the observed Ti/Fe ratio is consistent with the solar ratio (Fig.~\ref{fig:f3}), whereas the chemical evolution of our Galaxy requires overproduction of Ti for typical CC SNe ([Ti/Fe]$\sim$ 0.4\cite{2006ApJ...653.1145K}). While the deficiency of titanium in the SN yield may come from atomic data/nuclear reaction rate uncertainty, another possibility is that it could be related to the yet-unknown explosion mechanism on which the present result may shed some light. The high Ti yield exceeding the solar ratio, necessary to reproduce the abundances of the metal-poor stars, may be realized if the hot and high-entropy bubble could significantly contribute the titanium production\cite{1999ApJ...511..341N,2020ApJ...895...82V}; the Ti/Fe is limited to the solar value for the incomplete (QSE) Si burning regime. It can exceed the solar value significantly for the high entropy bubble (Fig.~\ref{fig:f3}), especially for the higher explosion energy (or higher entropy) and low $Y_{\rm e}$ below 0.5.

To realize [Ti/Fe]$\sim$ 0.4 (Ti/Fe mass ratio $\sim$ 0.45\%) in a SN CC, most of the iron-rich ejecta needs to be synthesized in a high-entropy environment. If we assumed the Ti/Fe mass ratios of 10$^{-3}$ and 10$^{-2}$ for normal Fe-rich ejecta and extremely high-entropy Fe-rich ejecta, respectively, 40\% of the Fe-rich ejecta must be synthesized at the extremely high-entropy enviroment. In addition, such a quite high Ti/Fe mass ratio of 10$^{-2}$ is hardly achieved with 1D SN calculations (Extended Data Fig.~\ref{fig:E5}). In the three-dimensional simulation of Vance et al. (2020)\cite{2020ApJ...895...82V}, the hottest ejecta have Ti/Fe $\sim$ 10$^{-2}$ where the thermodynamic trajectories from the three-dimensional supernova explosion model enhanced the Ti production (although this is radioactive $^{44}$Ti). If there is a SN CC whose most of the iron-rich ejecta are synthesized in such an environment, the overabundance of Ti in the early phase might be explained by it. Thus, future multi-dimensional simulations will help us to understand the Galactic chemical evolution, too.

\vspace{1cm}
\noindent
{\bf Data availability}~ All the {\it Chandra} and {\it NuSTAR} data using this research are available from the {\it Chandra} Data Archive (CDA: \url{https://cxc.harvard.edu/cda/}) and the {\it NuSTAR} Archive (\url{https://heasarc.gsfc.nasa.gov/docs/nustar/nustar\_archive.html}) in raw and reduced formats.

\noindent
{\bf Code availability}~ To analyze X-ray data with {\it Chandra}, we used public software, Chandra Interactive Analysis of Observations: CIAO (\url{https://cxc.cfa.harvard.edu/ciao/}). We used public atomic data in atomDB \url{(http://www.atomdb.org/}) and SPEX (\url{https://www.sron.nl/astrophysics-spex}). We fitted the X-ray spectra with a public package, Xspec (\url{https://heasarc.gsfc.nasa.gov/xanadu/xspec/}). We have not made publicly available codes for the hydrodynamics and nucleosynthesis of supernova explosions because they are not prepared for the open-use. Instead, the simulated thermodynamic profiles of the supernova explosions and the composition distributions shown in this paper are available on request.

\noindent
{\bf Acknowledgements}~ TS was supported by the Japan Society for the Promotion of Science (JSPS) KAKENHI grant No. JP19K14739, the Special Postdoctoral Researchers Program, and FY 2019 Incentive Research Projects in RIKEN. KM was supported in part by the Grants-in-Aid for the Scientific Research of Japan Society for the Promotion of Science (JSPS, No. JP18H05223 and JP20H00174). SN is partially supported by the Grants-in-Aid for “Scientific Research of JSPS (KAKENHI) (A) 19H00693,” Program of RIKEN for Evolution of Matter in the Universe (r-EMU), and Theoretical and Mathematical Sciences Program of RIKEN (iTHEMS). JPH acknowledges support for X-ray studies of SNRs from NASA grant NNX15AK71G to Rutgers University. TY is supported in part by the Grants-in-Aid for Scientific Research of Innovative Areas (JP20H05249). HU is supported in part by the Grant-in-Aid for Scientific Research (JP17H01130).

\noindent
{\bf Author contributions}~ TS wrote the manuscript with comments from all the authors and analysed the {\it Chandra} data. KM, SN, HU, JPH and BJW made significant contributions to the overall science case and manuscript. BG analysed the {\it NuSTAR} data and made Fig.~\ref{fig:f1}. TY, HU and MO calculated the nucleosynthesis models.

\noindent
{\bf Author Information}~ The authors declare no competing financial interests. All correspondence should be addressed to TS (toshiki.sato@riken.jp)

\newpage

\renewcommand{\figurename}{\bf Extended Data Fig.}
\setcounter{figure}{0}

%%%%%%%%%%%%%%%%%%%%%%%%%%%%%%%%%%%%%%%%%%%%%%%%%%%%%%%%%%%%%%%%%%%%%%%%%%%%%%%

\begin{figure}
\internallinenumbers % required by nature
\begin{center}
\includegraphics[width=14cm, bb=0 0 748 1000]{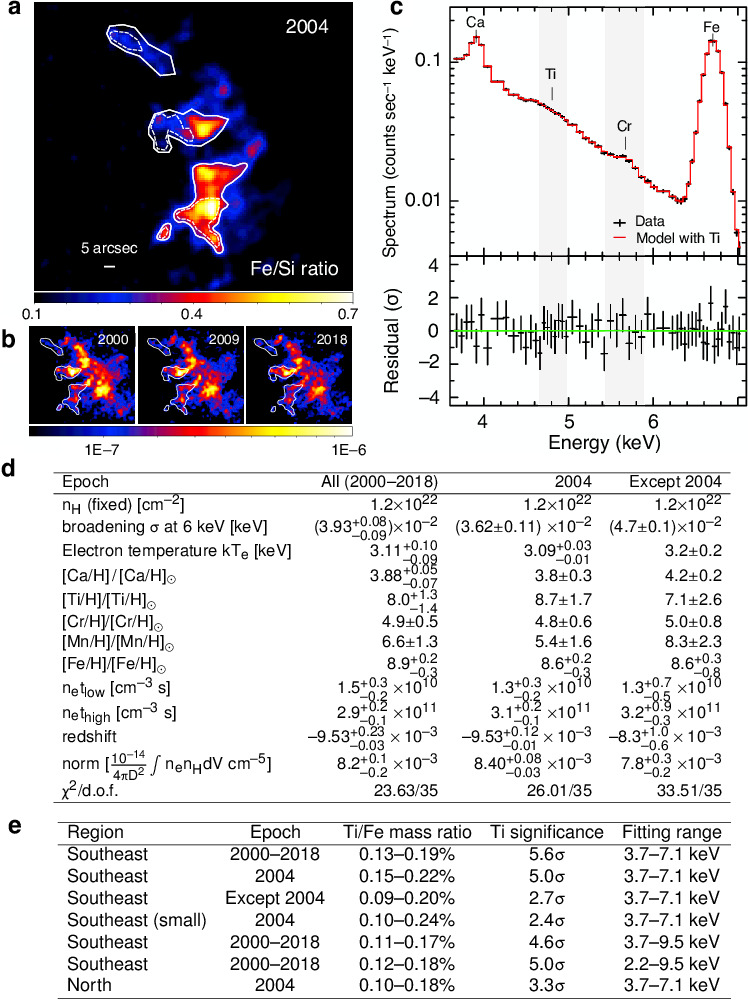}
\end{center}
\caption{%\scriptsize
\textbf{X-ray analysis for the southeastern iron-rich region.} {\bf a.} The Fe-K/Si-K ratio map in 2004. The solid white contour shows the region used for the result in the main text. {\bf b.} The Fe-K image in 2000 (left), 2009 (middle) and 2018 (right). In order to track the proper motions of each structure, we shift the region from epoch to epoch. {\bf c.} The X-ray spectrum and its best-fit model for the southeastern iron-rich region. The spectrum (black data) taken is the same as in Fig.~\ref{fig:f2}, but the best-fit thermal model has the Ti and Cr emissions. The residuals around 4.7--4.8 keV and 5.5--5.9 keV in Fig.~\ref{fig:f2} are well explained by the Ti and Cr emissions. {\bf d.} The best-fit parameters for the iron-rich ejecta. The errors show 1$\sigma$ confidence level ($\Delta\chi^2$ = 1.0). The solar abundance in Anders E. \& Grevesse N. (1989)\cite{1989GeCoA..53..197A} is used. {\bf e.} The summary of the Ti measurements in the Fe-rich regions of Cassiopeia A. The errors show 1$\sigma$ confidence level ($\Delta\chi^2$ = 1.0).}
\label{fig:E1}
\end{figure}

%%%%%%%%%%%%%%%%%%%%%%%%%%%%%%%%%%%%%%%%%%%%%%%%%%%%%%%%%%%%%%%%%%%%%%%%%%%%%%%
\begin{figure}
\internallinenumbers % required by nature
\begin{center}
\includegraphics[width=13.5cm, bb=0 0 849 1097]{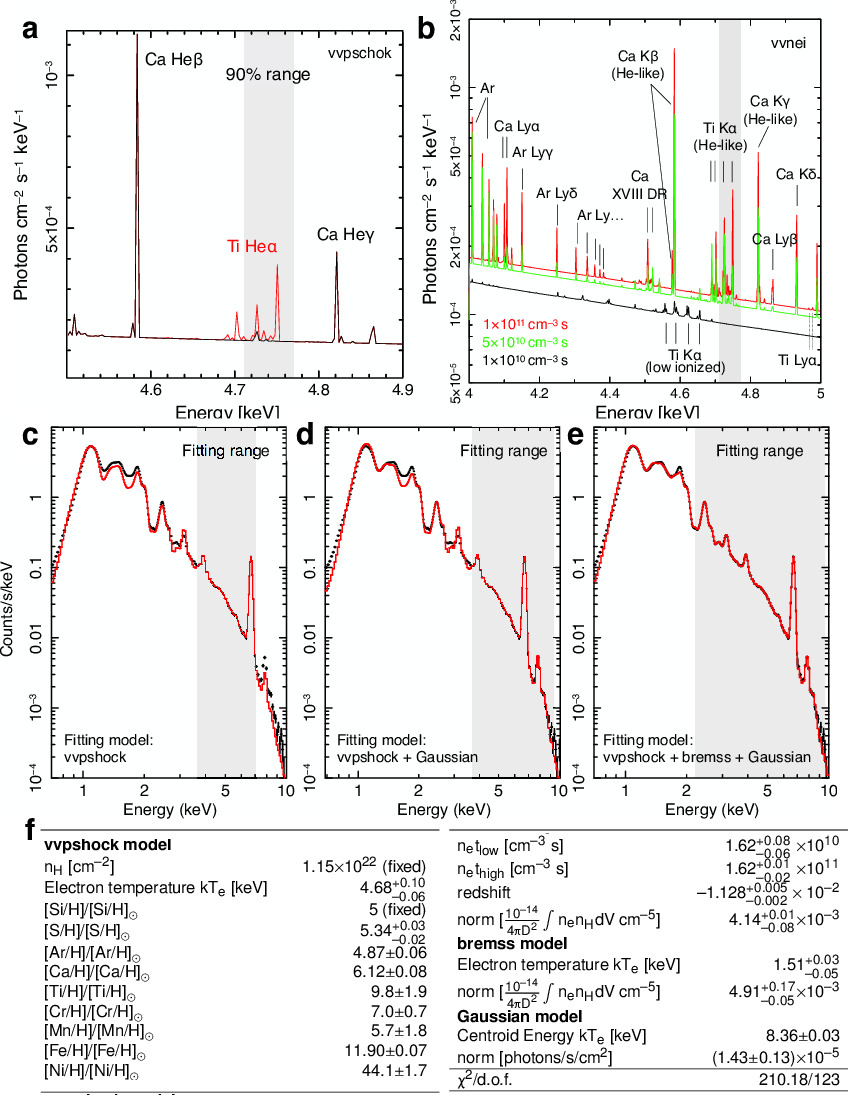}
\end{center}
\caption{%\scriptsize
\textbf{X-ray spectral modeling around the Ti line.}
\textbf{a.} The red and black show a plasma model (vvpshock) with titanium and without titanium, respectively. The He$\alpha$ emissions from titanium (red) are between Ca He$\beta$ and Ca He$\gamma$. The gray area shows the 90\% error range of the centroid energy of the Ti line by {\it Chandra}. The plasma parameters are the same as in Extended Data Fig.~\ref{fig:E1}e (without the line broadening). \textbf{b.} Comparison of 4--5 keV model spectra (vvnei) that have different ionization states. The gray area shows the 90\% error range of the centroid energy of the Ti line by {\it Chandra}. The most prominent lines are the Ca He$\beta$ and Ca He$\gamma$ lines at 4.584 keV and 4.822 keV, respectively. These two Ca lines become stronger at high ionization states ($>5\times10^{10}$ cm$^{-3}$ s). \textbf{c.} The black data and red curves show the observed spectra and the best-fit models, respectively. The fitting range (gray area) is 3.7--7.1 keV. This result is used in the main text. \textbf{d.} The fitting range is 3.7--9.5 keV where the emissions up to Ni are included. To express the lack of emissions around 8.3 keV, one Gaussian line is added. \textbf{e.} The fitting range is 2.2--9.5 keV. Here, we added a thermal bremsstrahlung model to express a low temperature component. The best-fit parameters are summarized in table {\bf f}. {\bf f.} The best-fit parameters for the Fe-rich ejecta in the spectrum {\bf e}. The errors show 1$\sigma$ confidence level ($\Delta\chi^2$ = 1.0). The solar abundance in Anders E. \& Grevesse N. (1989)\cite{1989GeCoA..53..197A} is used. Some other lighter elements that are not shown here are also included in the model.
}
\label{fig:E2}
\end{figure}

%%%%%%%%%%%%%%%%%%%%%%%%%%%%%%%%%%%%%%%%%%%%%%%%%%%%%%%%%%%%%%%%%%%%%%%%%%%%%%%

\begin{figure}
\internallinenumbers % required by nature
\resizebox{\hsize}{!}{\includegraphics[bb=0 0 1642 1497]{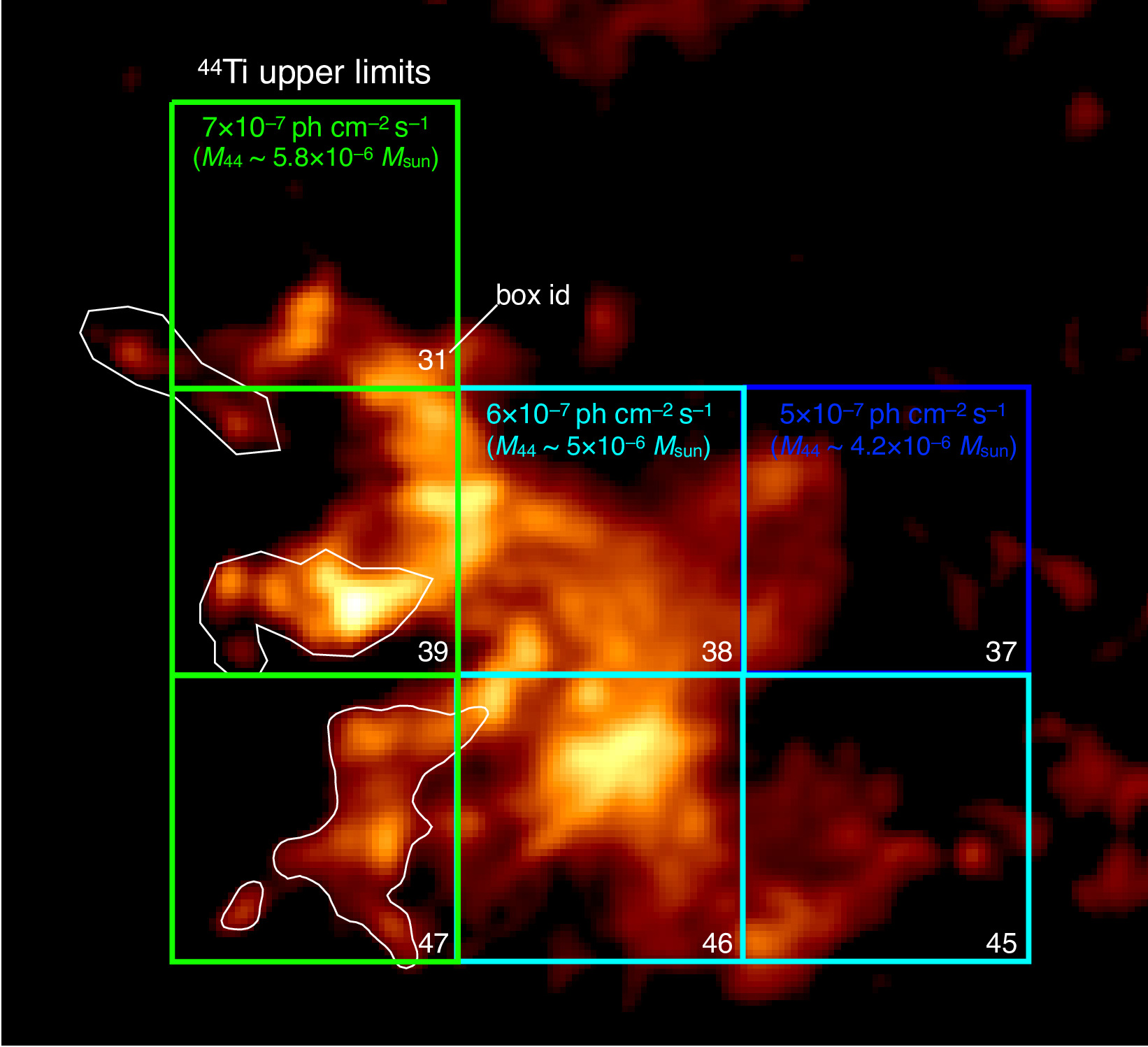}}
\caption{%\scriptsize
\textbf{The iron distribution (image) and the $^{44}$Ti upper limit map (colored boxes) around the southeastern region.}
We use the $^{44}$Ti upper limits estimated in Grefenstette et al. (2017)\cite{2017ApJ...834...19G}. The box size is 45$^{\prime\prime}$ $\times$ 45$^{\prime\prime}$. The box ids are the same as those in the paper. The white contour regions are the same as Fig.~\ref{fig:f1}. Almost all of the areas we extracted spectrum are included in three box regions: 31, 39 and 47.}
\label{fig:E3}
\end{figure}

%%%%%%%%%%%%%%%%%%%%%%%%%%%%%%%%%%%%%%%%%%%%%%%%%%%%%%%%%%%%%%%%%%%%%%%%%%%%%%%

\begin{figure}
\internallinenumbers % required by nature
\resizebox{\hsize}{!}{\includegraphics[bb=0 0 2448 1868]{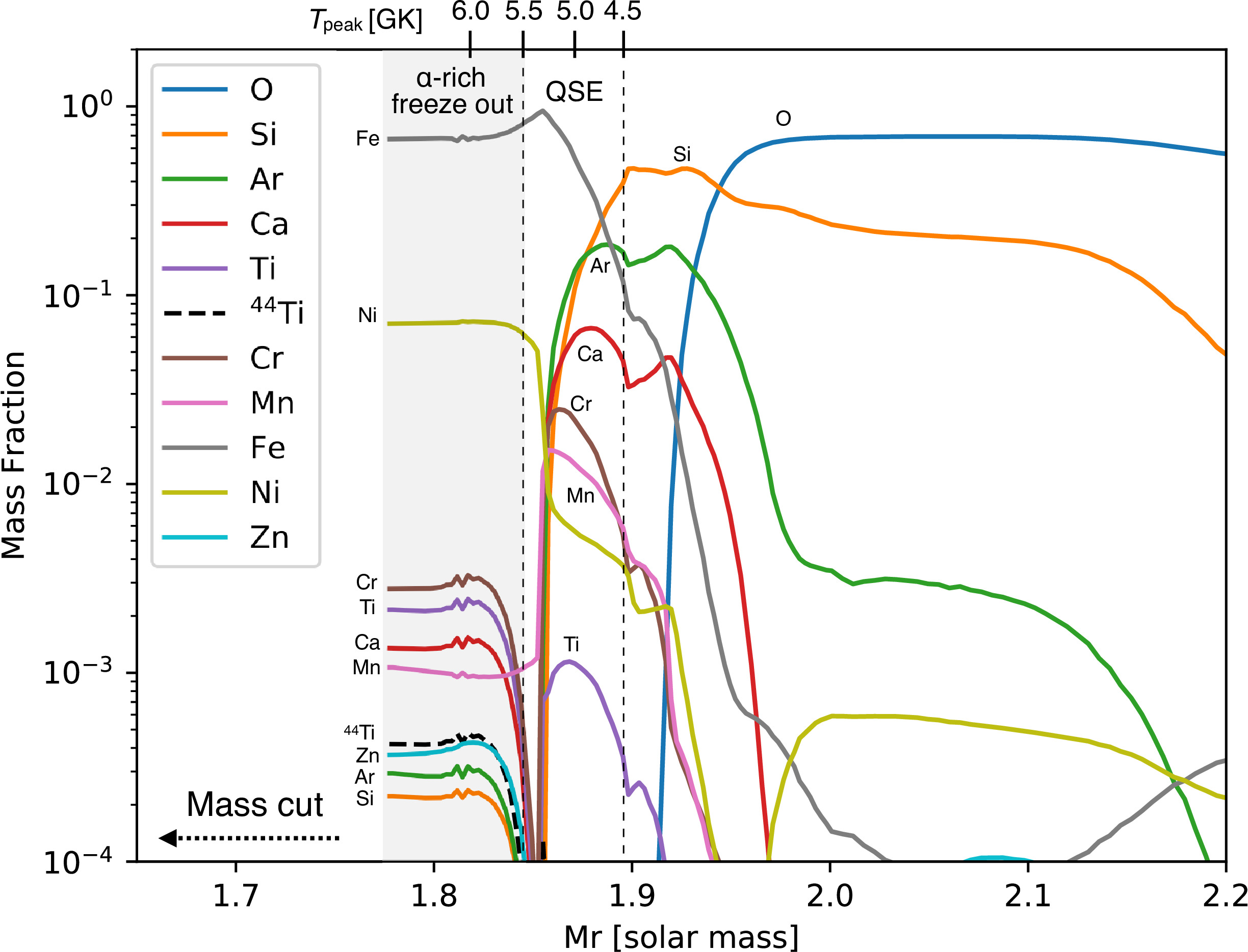}}
\caption{%\scriptsize
\textbf{The one-dimensional CC SN nucleosynthesis model used in this study.}
The model assumed a high energy explosion energy of 3$\times$10$^{51}$ erg for a 15 $M_{\odot}$ progenitor. The $\alpha$-rich freeze out produces some $\alpha$ elements (e.g., Fe, Ni, Cr, Ti, Zn) at the deepest layer with high peak temperatures ($>$5.5 GK). At the QSE (i.e., incomplete Si burning) layer, the intermediate mass elements (e.g., Si, S, Ar, Ca, Cr, Mn) are abundant.}
\label{fig:E4}
\end{figure}

%%%%%%%%%%%%%%%%%%%%%%%%%%%%%%%%%%%%%%%%%%%%%%%%%%%%%%%%%%%%%%%%%%%%%%%%%%%%%%%

%%%%%%%%%%%%%%%%%%%%%%%%%%%%%%%%%%%%%%%%%%%%%%%%%%%%%%%%%%%%%%%%%%%%%%%%%%%%%%%

\begin{figure}
\internallinenumbers % required by nature
\begin{center}
\includegraphics[bb=0 0 977 1162, width=16cm]{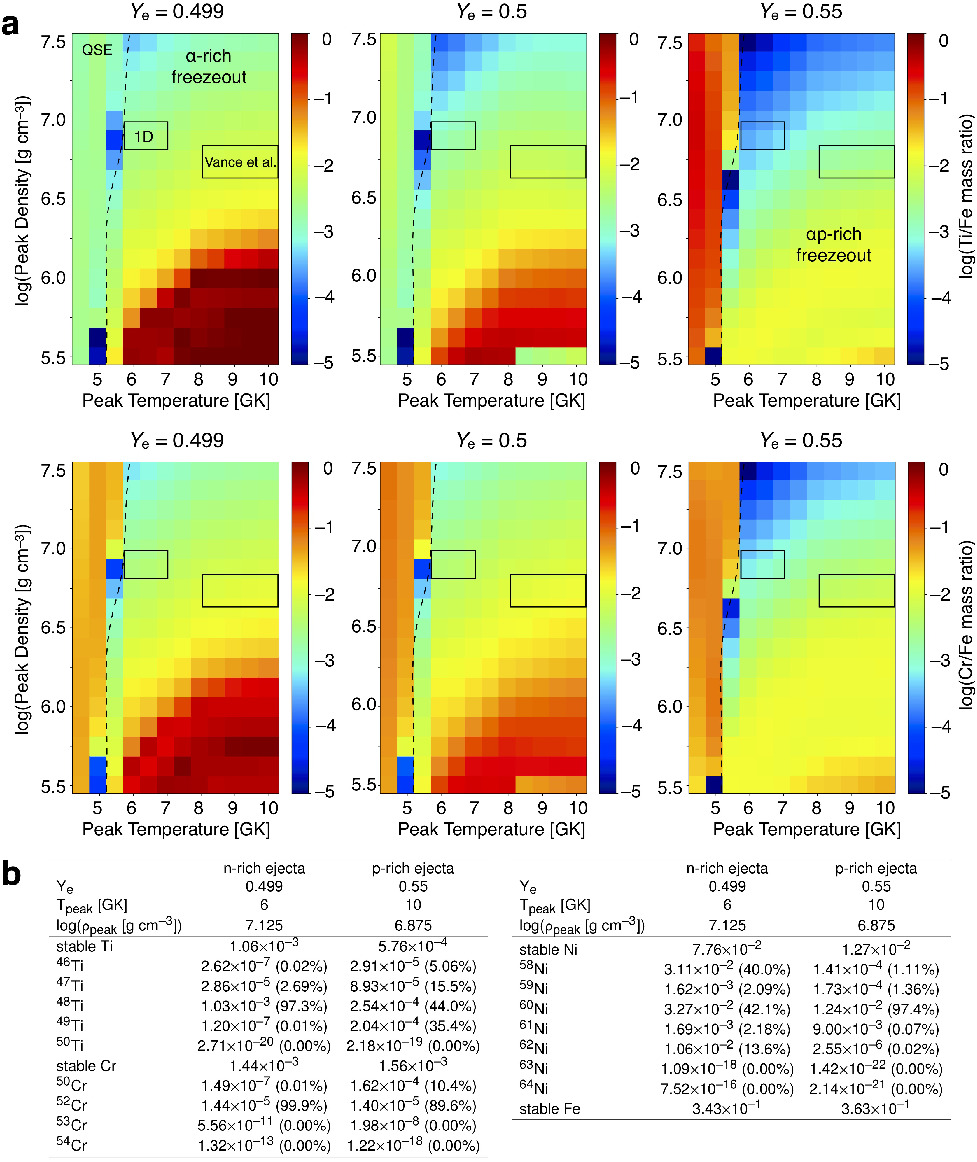}
\end{center}
\caption{%\scriptsize
\textbf{Nucleosynthesis calculations in the peak temperature–density plane.} {\bf a.} Ti/Fe (top row) and Cr/Fe (bottom row) mass ratios in the peak temperature–density plane. From left to right, the lepton fraction corresponds to $Y_{\rm e}$ = 0.499, 0.5, and 0.55. Here, we used the thermodynamic trajectories taken from our 1D SN model. All the stable isotopes are included. The production of Ti and Cr is sensitive to the high-entropy environment (toward the bottom right), which is the same as radioactive $^{44}$Ti\cite{2010ApJS..191...66M}. The broken lines show the boundary between incomplete and complete Si burning. The black boxes show typical parameter spaces for the complete Si burning ($\alpha$-rich freezeout) in our 1D SN model and the 3D SN model in Vance et al. (2020)\cite{2020ApJ...895...82V}. {\bf b.} Mass fractions of Ti, Cr and Ni isotopes in the nucleosynthesis calculations.
}
\label{fig:E5}
\end{figure}

%%%%%%%%%%%%%%%%%%%%%%%%%%%%%%%%%%%%%%%%%%%%%%%%%%%%%%%%%%%%%%%%%%%%%%%%%%%%%%%

\begin{figure}
\internallinenumbers % required by nature
\resizebox{\hsize}{!}{\includegraphics[bb=0 0 1628 1233]{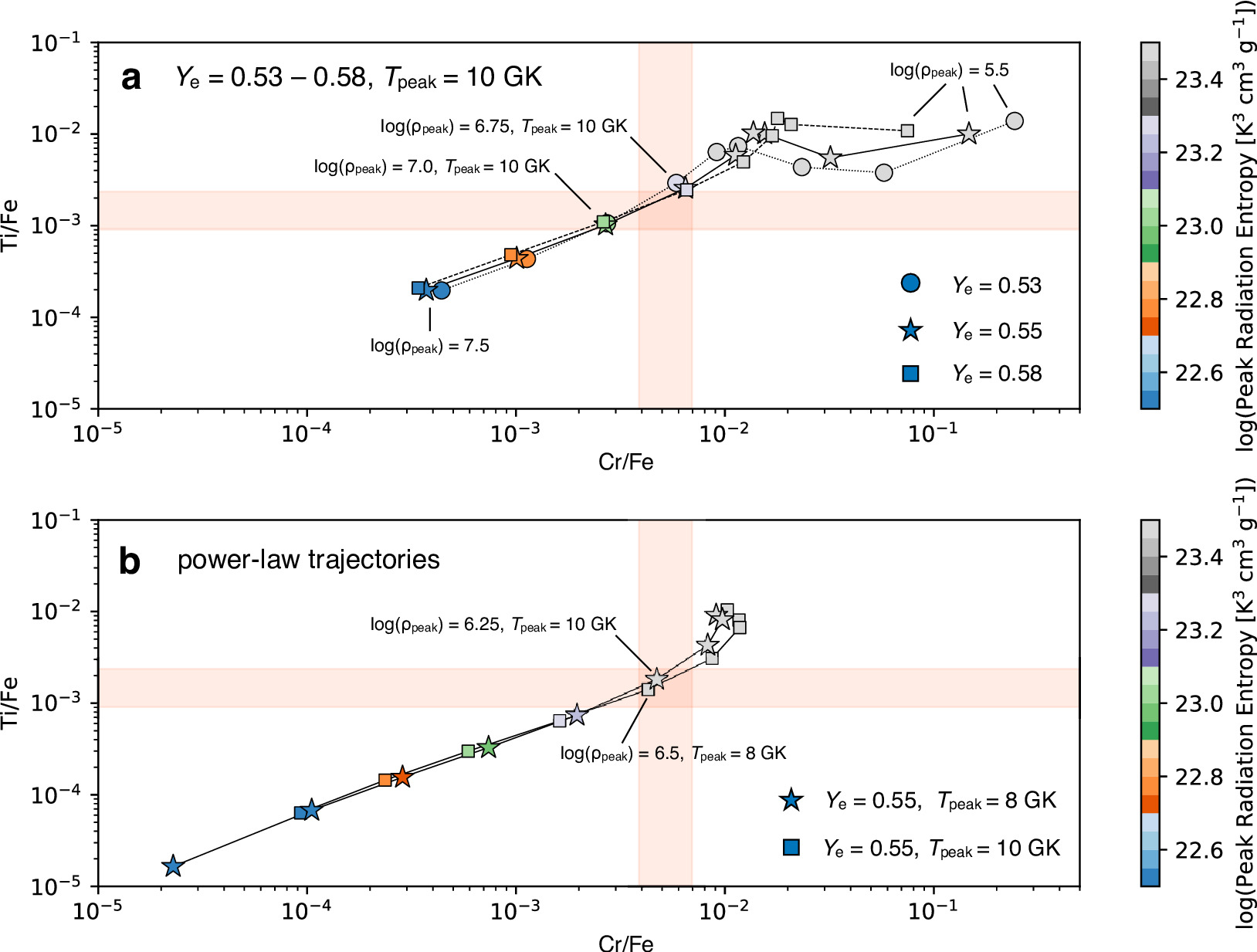}}
\caption{%\scriptsize
\textbf{The observed Ti/Fe and Cr/Fe mass ratios and nucleosynthesis models.} The colored areas show the observed mass ratios.
\textbf{a.} The colored points show parameter studies for hot ($T_{\rm peak} =$ 10 GK) and proton-rich environment while changing the peak density from 10$^{5.5}$ g cm$^{-3}$ to 10$^{7.5}$ g cm$^{-3}$. The circle, star and box show $Y_{\rm e}$ = 0.53, 0.55 and 0.58, respectively. Here, we used the thermodynamic trajectories taken from our 1D SN model. \textbf{b.} Parameter studies with power-law thermodynamic evolusion. The star and box data show the Ti/Fe and Cr/Fe mass ratios produced by $T_{\rm peak} =$ 8 GK and $T_{\rm peak} =$ 10 GK, respectively. To reproduce the observed mass ratios, higher radiation entropy than that in the model with the thermodynamic trajectories taken from the 1D SN model is needed.
}
\label{fig:E6}
\end{figure}

%%%%%%%%%%%%%%%%%%%%%%%%%%%%%%%%%%%%%%%%%%%%%%%%%%%%%%%%%%%%%%%%%%%%%%%%%%%%%%%

\begin{figure}
\internallinenumbers % required by nature
\resizebox{\hsize}{!}{\includegraphics[bb=0 0 1132 845]{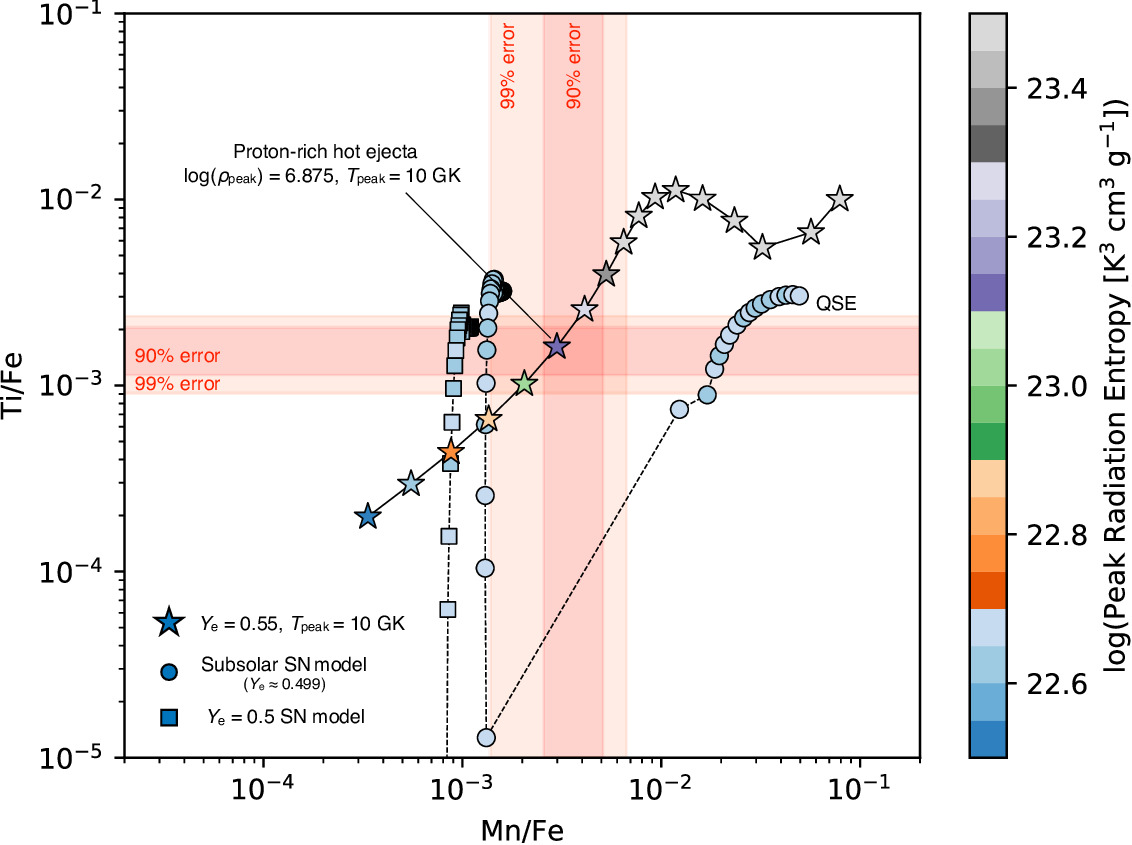}}
\caption{%\scriptsize
\textbf{The observed Ti/Fe and Mn/Fe mass ratios and nucleosynthesis models.}
The Mn/Fe mass ratio is derived from the best-fit model in the left column of Extended Data Fig.~\ref{fig:E1}e. The colored areas show the observed mass ratios with 90\% and 99\% error range. The colored circles show the mass ratios in our 1D SN model ($M_{\rm ZAMS} =$ 15 $M_{\odot}$, $E_{\rm exp} = 3\times10^{51}$ erg, $Z = 0.5 Z_\odot$). In the box data, the same 1D SN model was used, but the lepton fraction at the $\alpha$-rich freeze out is modified to $Y_{\rm e} = 0.5$. The modification from circle to box (increase of $Y_{\rm e}$) suppresses the synthesized amount of neutron-rich elements like Mn. The colored stars show a parameter study for hot ($T_{\rm peak} =$ 10 GK) and proton-rich ($Y_{\rm e}$ = 0.55) environment while changing the peak density from 10$^{5.5}$ g cm$^{-3}$ to 10$^{7.5}$ g cm$^{-3}$.
}
\label{fig:E7}
\end{figure}

%%%%%%%%%%%%%%%%%%%%%%%%%%%%%%%%%%%%%%%%%%%%%%%%%%%%%%%%%%%%%%%%%%%%%%%%%%%%%%%

\begin{figure}[h]
\internallinenumbers % required by nature
\resizebox{\hsize}{!}{\includegraphics[bb=0 0 780 701]{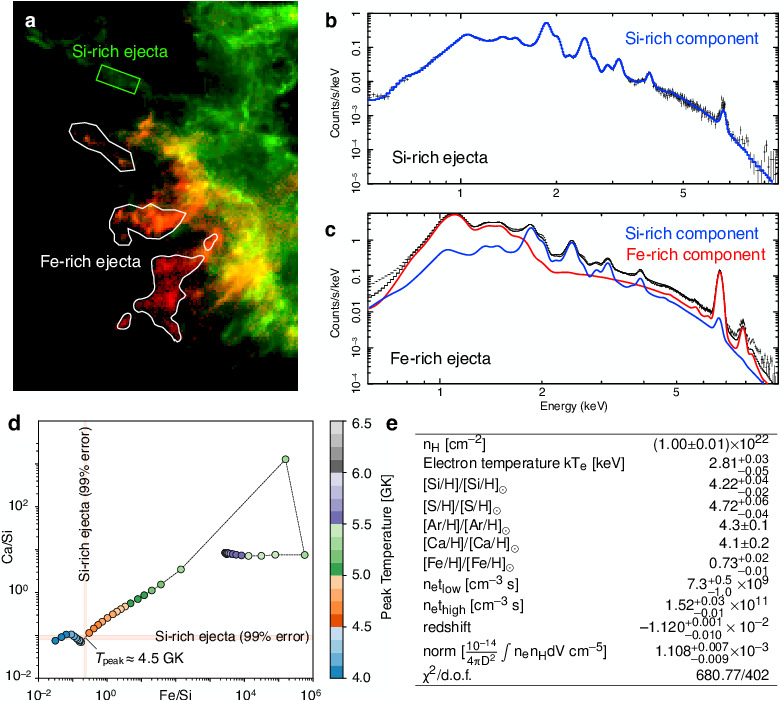}}
\caption{%\scriptsize
\textbf{Comparison between the Fe-rich and Si-rich regions.}
{\bf a.} Two color image around the southeastern iron-rich region. The red and green show the Fe and Si images, respectively. The green box is defined as the Si-rich ejecta region. {\bf b.} The X-ray spectrum extracted from the Si-rich region. The blue curve (Si-rich component) shows the best-fit thermal model. {\bf c.} The X-ray spectrum at the Fe-rich region. The Si-rich  component has the same plasma parameters as in the model of {\bf b} The red model shows the Fe-rich component that has emissions from H, He, Ti, Cr, Mn, Fe and Ni. {\bf d.} Comparisons of the observed Ca/Si and Fe/Si mass ratios in the Si-rich ejecta region with those by theoretical calculations.
The faint orange areas are the observed mass ratios (99\% confidence level,$\Delta\chi^2$= 6.64).  The colored points show the mass ratios of the nucleosynthesis calculations in Fig.~\ref{fig:f3}. {\bf e.} The best-fit parameters for the Si-rich ejecta in the spectrum {\bf b}. The spectrum is extracted from the 2004 data. The errors show 1$\sigma$ confidence level ($\Delta\chi^2$ = 1.0). The solar abundance in Anders E. \& Grevesse N. (1989)\cite{1989GeCoA..53..197A} is used.}
\label{fig:E8}
\end{figure}

%%%%%%%%%%%%%%%%%%%%%%%%%%%%%%%%%%%%%%%%%%%%%%%%%%%%%%%%%%%%%%%%%%%%%%%%%%%%%%%

\begin{figure}[h]
\internallinenumbers % required by nature
\resizebox{\hsize}{!}{\includegraphics[bb=0 0 2516 1276]{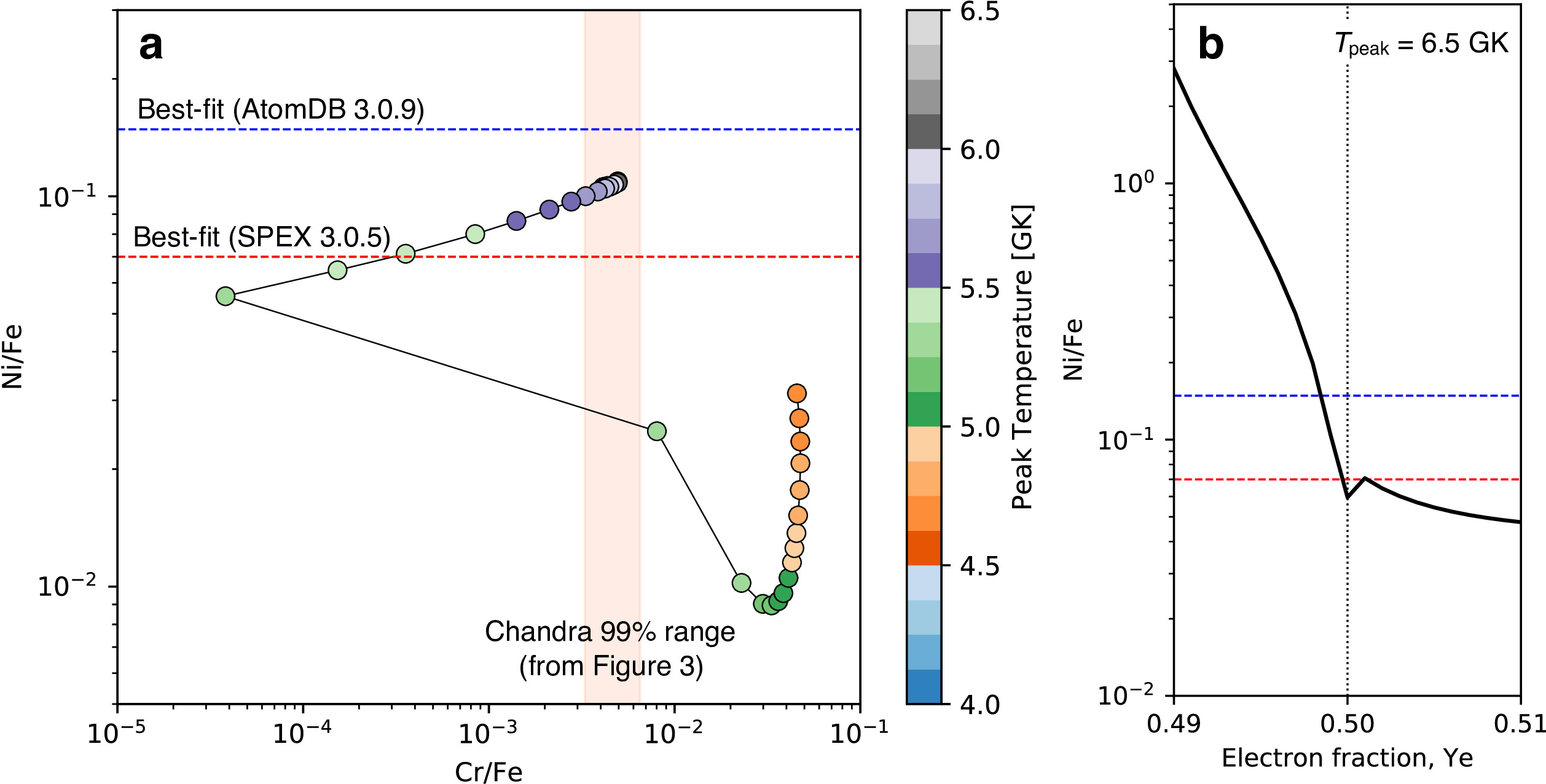}}
\caption{%\scriptsize
\textbf{Comparisons of the observed Ni/Fe and Cr/Fe mass ratios in the Fe-rich ejecta region with those by theoretical calculations.}
\textbf{a.} The blue and red broken lines show the best-fit Ni/Fe mass ratios with Xspec (AtomDB\cite{2012ApJ...756..128F}, version 3.0.9) and SPEX\cite{1996uxsa.conf..411K} (version 3.0.5), respectively. The best-fit Ni/Fe mass ratios are different from each other because the emissivities of the Fe K$\beta$,$\gamma$,$\delta$,... emissions are different from atomic code to atomic code. The faint orange areas are the observed Cr/Fe mass ratios in Fig.~\ref{fig:f3}.  The colored points show the mass ratios of the nucleosynthesis calculations in Fig.~\ref{fig:f3}.
\textbf{b.} The Ni/Fe dependence on the lepton (electron) fraction, $Y_{\rm e}$. The blue and red broken lines show the best-fit Ni/Fe mass ratios with Xspec  and SPEX, respectively. Here, we assumed the explosion energy of 3$\times$10$^{51}$ erg, and a region with the peak temperature of $T_{\rm peak}$ = 6.5 GK is analyzed. In the neutron-rich side, the Ni/Fe ratio changes more dramatically because the neutron-rich element, $^{58}$Ni is efficiently synthesized. On the other hand, in the proton-rich side, the Ni is not so sensitive to the lepton fraction. Here, $^{60}$Ni is dominantly synthesized (see Extended Data Fig.~\ref{fig:E5}b).
}
\label{fig:E9}
\end{figure}

\begin{figure}[h]
\internallinenumbers % required by nature
\resizebox{\hsize}{!}{\includegraphics[bb=0 0 1849 1549]{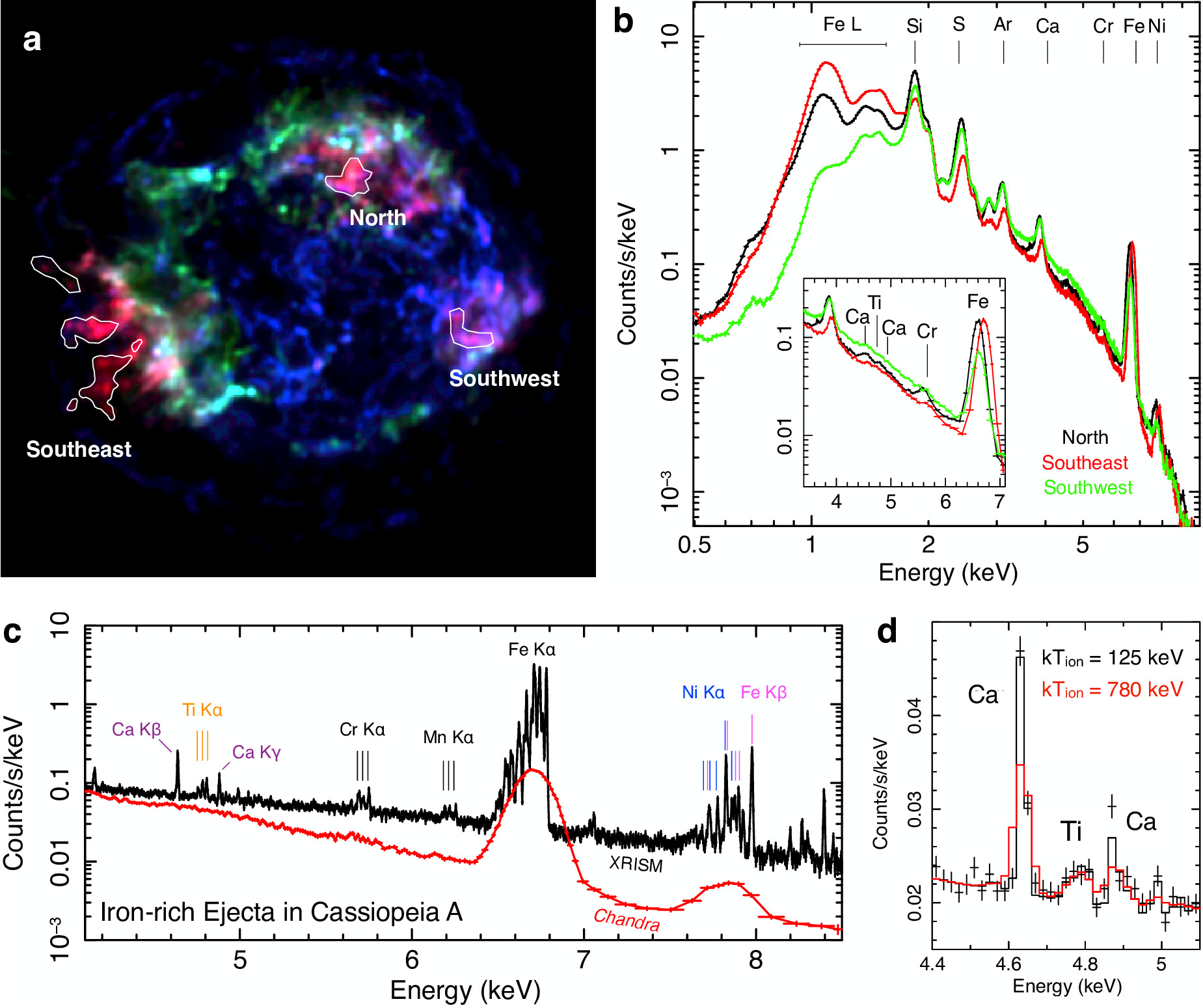}}
\caption{%\scriptsize
\textbf{Comparison of three Fe-rich regions in Cassiopeia A.}
\textbf{a.} three-color image of Cassiopeia A. The red, green and blue show the Fe-K, Si-K and continuum ($\approx$non-thermal) emissions, respectively. \textbf{b.} X-ray spectra in the southeast (red), north (black) and southwest (green) regions. \textbf{c.} Comparison of spectra between XRISM and Chandra. We assumed an energy resolution of 7 eV (FWHM) and exposure time of 1 Msec for the XRISM simulation. In the simulated XRISM spectrum, we do not consider the line broadening effects (either thermal and Doppler). \textbf{d.} Zoom up around the Ti emissions. Here we simulated a spectrum with the thermal broadening assuming $kT_{\rm ion}$ = 125 keV (data with error bars). The black and red lines show the thermal models with $kT_{\rm ion}$ = 125 keV and $kT_{\rm ion}$ = 780 keV, respectively.}
\label{fig:E10}
\end{figure}

%%%%%%%%%%%%%%%%%%%%%%%%%%%%%%%%%%%%%%%%%%%%%%%%%%%%%%%%%%%%%%%%%%%%%%%%%%%%%%%

\clearpage
%\noindent{\large\bf Bibliography}%\vspace{1.cm}

\end{document}